\documentclass[a4paper,10pt,twoside]{cpc-hepnp}
\usepackage{multicol}
\usepackage{graphicx}
\usepackage{booktabs}
\usepackage{amssymb,bm,mathrsfs,bbm,amscd}
\usepackage[tbtags]{amsmath}
\usepackage{lastpage}
\usepackage{CJK}
\usepackage{caption}
\usepackage{geometry}
\begin{document}
\begin{CJK*}{GBK}{song}

\fancyhead[c]{\small Chinese Physics C~~~Vol. XX, No. X (2019)
XXXXXX} \fancyfoot[C]{\small XXXXXX-\thepage}

\title{The expectation of cosmic ray proton and helium energy spectrum below 4 PeV measured by LHAASO}

\author{%
     L.Q. Yin$^{1,2}$\email{yinlq@ihep.ac.cn}%
\quad S.S. Zhang$^{1}$,
Z. Cao$^{1,2}$,
B.Y. Bi$^{3}$,
C. Wang$^{1}$,
J.L. Liu$^{4}$,
L.L. Ma$^{1}$,\\
M.J. Yang$^{1}$,
Tiina Suomij\"{a}rvi$^{5}$,
Y. Zhang$^{1}$,
Z.Y. You$^{1,2}$,
Z.Z Zong$^{6}$\\
for the LHAASO Collaboration
}
\maketitle

\address{%

$^1$ Key Laboratory of Particle Astrophysics, Institute of High Energy Physics, Chinese Academy of Sciences,
100049 Beijing, China.\\
$^2$ University of Chinese Academy of Science, Beijing, China.\\
$^3$ Institut f\"{u}r Astronomie und Astrophysik, Eberhard Karls Universit\"{a}t, T\"{u}bingen 72076, Germany.\\
$^4$ Kunming University, Kunming, Yunnan, China.\\
$^5$ Institut de Physique Nucl\'{e}aire d'Orsay, IN2P3-CNRS, Universit\'{e} Paris-Sud, Universit\'{e}, Paris-Saclay, 91406 Orsay Cedex, France.\\
$^6$ China Nuclear Power Engineering Co., Ltd., 100084, Beijing, China
}

\begin{abstract}
Large High Altitude Air Shower Observatory(LHAASO) is a composite cosmic ray observatory consisting of three detector arrays:  kilometer square array (KM2A) which includes the electromagnetic detector array and muon detector array, water Cherenkov detector array (WCDA) and wide field of view Cherenkov telescope array (WFCTA).
One of the main scientific objectives of LHAASO is to precisely measure the cosmic rays energy spectrum of individual components from $10^{14}$ eV to $10^{18}$ eV.
The hybrid observation will be employed  by LHAASO experiment, in which the lateral and longitudinal distributions of the extensive air shower can be observed  simultaneously.
Thus many kinds of parameters can be used for primary nuclei identification.
In this paper, high purity cosmic ray simulation samples of light nuclei component are obtained through Multi-Variable Analysis.
The apertures of 1/4 LHAASO array for pure proton and mixed proton and helium (H\&He) samples are 
$900 \ m^{2}Sr$ and $1800 \ m^{2}Sr$ respectively.
A prospect of proton and H\&He spectra from 100 TeV to 4 PeV is discussed.

\end{abstract}


\section{Introduction}

An unsolved problem in cosmic ray observation is the ``knee" structure in the energy spectrum, 
namely a significant bending of the spectrum from the power-law index of approximately -2.7 to -3.1 around a few PeV.
The causes of the knee structure are closely related to the origin, acceleration and propagation mechanism of cosmic rays, but there is no consistent measurement result so far.
Several experiments have measured the all-particle spectra around the knee region, such as KASCADE-Grand~\cite{kascade-grand}, Tibet-AS$\gamma$~\cite{ASr}, EAS-TOP~\cite{eas-top}, etc.
These experiments exhibited a similar knee-like structure at energies of about 4 PeV, 
within a factor of two in the flux values~\cite{knee_summary}.

In addition, the energy spectra of individual components are much different.
For example, the unfolded proton spectrum measured by the KASCADE experiment shows the steepening at 2 PeV or 3PeV based on QGSJET01 model and SIBYLL2.1 model respectively~\cite{kascade}. 
However, the hybrid experiment of ARGO-YBJ and two prototypes of WFCTA shows that the energy spectrum for H\&He has a knee-like structure at 700 TeV~\cite{argo-wfcta}.
The main reasons for this situation are the lack of absolute energy scale and the way to identify the type of the primary particles of cosmic rays. 
LHAASO~\cite{lhaaso1, lhaaso2}, located in Daocheng Haizishan, 4410 m a.s.l., Sichuan Province, China, will throw light on this traditional problem.

The LHAASO site is at an ideal altitude for knee physics research because the atmospheric depth is close to the maximum of the development of cosmic ray extensive air shower (EAS)~\cite{EAS},
so that the fluctuation of EAS and the dependence of the interaction model will be the minimum.
Similar to the ARGO-YBJ experiment, WCDA can obtain absolute-energy-scale through moon shadow observation~\cite{argo-moon}.
Because the detection threshold of WCDA can be as low as several hundred GeV, 
it can be directly compared with the measurement results of space experiments 
to study the error of energy scale.

Furthermore, LHAASO contains four types of detectors, which can detect electromagnetic particles, muons, Cherenkov and fluorescent photons in the EAS.
Multi-Variable measurements of the EAS can effectively distinguish the components of cosmic rays.
Thus, the energy spectra of the individual components can be precisely measured by LHAASO through Multi-Variable Analysis (MVA).

LHAASO is planed to obtain the consecutive measurement of energy spectra by four stages.
First of all, WCDA delivers the absolute-energy-scale to WFCTA  
by hybrid observation of cosmic rays from 10 TeV to 100 TeV.
In this energy band, the energy spectra measured by LHAASO can overlap with the spectra of space experiments.
The second stage is mainly for the knee observation of light nuclei components of cosmic rays in the energies from 100 TeV to 10 PeV by the hybrid detection of WFCTA, WCDA, and KM2A.
In the third stage, the layout of WFCTA will be changed to measure the knee of the heavy nuclei from 10 PeV to 100 PeV~\cite{llma:iron_knee}.
WCDA will not be included in the hybrid detection due to saturation.
The last stage is for the second knee study from 100 PeV to 1 EeV, 
in which WFCTA will be operated in fluorescence observation mode~\cite{liujiali:fluorescence}.

In the first three observation stages, the elevation angles of the WFCTA are 90$^{o}$, 60$^{o}$, 45$^{o}$ respectively.
Their corresponding atmospheric depths are respectively around the shower maximum positions of cosmic rays in their own observation stages.
This paper is devoted to the prospect of energy spectra observation of pure proton and H\&He in the second stage through simulation.

\section{The Hybrid Experiment}\label{section:experiment}

\begin{figure}[htb]
	\centering
	\includegraphics[width=0.6\linewidth]{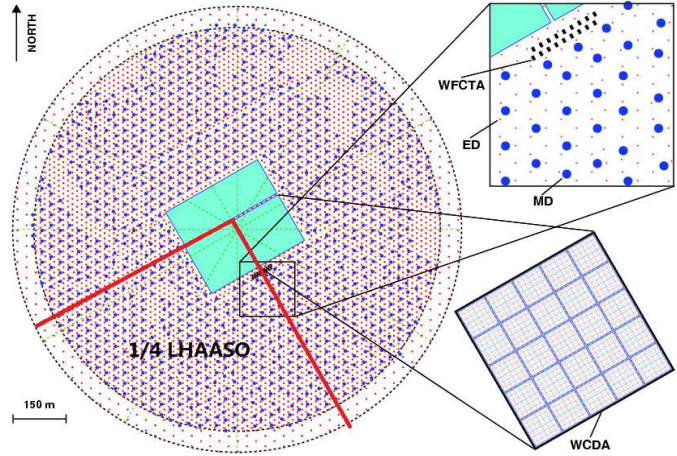}
	\caption{The detectors layout of LHAASO experiment.}
	\label{lhaaso_array}
\end{figure}

LHAASO was formally approved on December 31, 2015 and now is under construction.
The detectors layout of LHAASO experiment is shown in Fig.~\ref{lhaaso_array}.
LHAASO consists of four types of detectors: electromagnetic detectors (red dots), muon detectors (blue dots),
water Cherenkov detector (blue squares) and wide field of view (FOV) Cherenkov telescopes(black rectangles).
There are many vacancies in the muon detectors (blue dots) distribution, 
which is due to the terrain making it impossible to install detectors.

WCDA~\cite{wcda}, located at the center of LHAASO, is a water Cherenkov detector array with a total area of 78,000 $m^2$.
It detects the Cherenkov light produced in water by cascade processes of secondary particles in EAS.
WCDA consists of three water ponds: two have area of $150 m\times150 m$ and one has area of $300 m\times110 m$.
All of the ponds have the water depth of 4.5 m.
Each pond is divided into small cells with an area of $5 m\times5 m$, 
having two Photomultiplier Tubes (PMT) anchored on the bottom at the center of the cell.
In one of the $150 m\times150 m$ ponds, the pond in the lower left quarter in Fig.~\ref{lhaaso_array},
a 1-inch PMT is placed right by an 8-inch PMT to enlarge the dynamic range to 200,000 photoelectrons.
It allows the measurement of cosmic rays with energy up to 10 PeV ~\cite{wcda++}.
This pond is called WCDA++ to achieve the hybrid detection with the WFCTA.

WFCTA is located at 85 meters away from the center of WCDA++.
It detects the Cherenkov and fluorescence photons in the air shower.
The angle of elevation of the telescope's main axis is set to $60^{o}$.
Each telescope includes a spherical mirror to collect Cherenkov and fluorescence photons and reflect them onto the camera~\cite{Camera} with the effective area of $5m^2$.
The camera is located at the focal point of the spherical mirror and its function is to convert Cherenkov/fluorescence photons into electrical signals.
The optical sensor of the camera is SiPM~\cite{SiPM}, with 1024 pieces in total.
The pixel of each SiPM is $0.5^{\circ}\times0.5^{\circ}$ and the camera watches a FOV of $16^{\circ}\times16^{\circ}$.
All the parts of the telescope are placed in a container for the convenience of movement and arrangement.
Two telescope prototypes have been operated successfully at Yangbajing cosmic ray observatory in Tibet~\cite{wfcta}.

KM2A contains two sub-arrays: electromagnetic detector array (ED) and muon detector array (MD).
ED array~\cite{ed} covers an area of 1.3 $km^{2}$ consisted of 5195 plastic scintillator detectors with a spacing of 15 meters. 
One ED detector has an effective area of 1 $m^{2}$.
As its name describes, the ED detects the electromagnetic particles in the EAS.
MD array is an underground water Cherenkov detector array~\cite{md}.
There are 1171 muon detectors also covering 1 $km^2$ area with a spacing of 30 meters.
The area of one muon detector is 36 $m^2$,
and it detects the information of muon in the EAS.

The 1/4 LHAASO array includes the WCDA++, six Cherenkov telescopes, 1272 EDs and 300 MDs.
The layout of the 1/4 array is shown in Fig.~\ref{lhaaso_array}.
This 1/4 array will run for a few years to achieve the physical goal of light nuclei components spectra measurement.

Since full-coverage array can provide more accurate geometric information of air showers, 
the secondary particles are mainly measured by WCDA++ in the second hybrid detection stage.
It means that the shower core position in this stage is outside of the ED array.
Therefore, ED is not included in this simulation.

\section{Simulation and Event Reconstructions}\label{section:simulation and reconstruction}

\subsection{Simulation}

\begin{figure}[htb]
	\centering
	\includegraphics[width=0.6\linewidth]{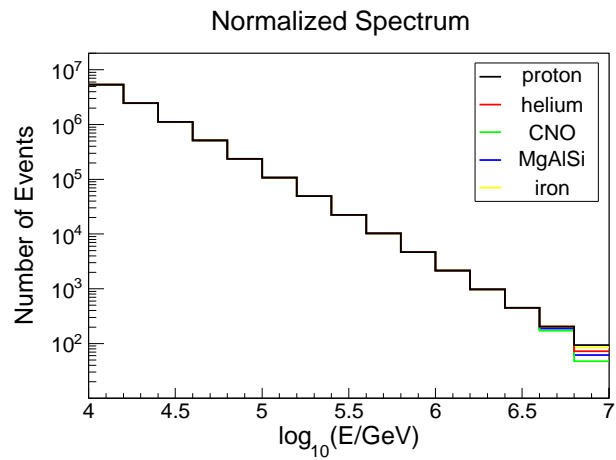}
	\caption{The energy distribution of all simulation events after normalization.}
	\label{raw_specra}
\end{figure}

The cascade processes of primary cosmic rays in the atmosphere are simulated by the CORSIKA~\cite{corsika} program  with the version of 6990.
The EGS4 model is chosen for the electromagnetic interaction.
The QGSJET02 and GHEISHA models are chosen for the high and low energy hadronic processes respectively.
Both the information of Cherenkov photons and secondary particle at the level of observatory are recorded to simulate the hybrid observation.
Five components, proton, helium, CNO, MgAlSi, and iron are generated with energies from 10 TeV to 10 PeV according to a power law spectrum with an index of -2.7.
The directions of the showers are from $24^{\circ}$ to $38^{\circ}$ in zenith and from $77^{\circ}$ to $103^{\circ}$ in azimuth.
The shower core position is evenly distributed in an area of 260 m$\times$260 m.
The primary energy spectrum is normalized to the exponent of -2.7 from 10 TeV to 10 PeV,
as shown in Fig.~\ref{raw_specra}.
Different components are normalized to the same with proton.

In addition, five components are also normalized according to H\"{o}randel model.
The relationship between the original energy spectra and the two normalized spectra is that 
the number of original proton events, 
the number of proton events with exponent of -2.7 and 
the number of proton events with exponent from H\"{o}randel model are the same at 100 TeV.
The events number of other energy bins and other components are normalized in proportion.
The normalized statistics of each component are equivalent to 
the observation data of one year with 10\% duty cycle
according to H\"{o}randel model~\cite{hrd}.

The simulation of the response of three LHAASO detector arrays for EAS is performed separately.
In the simulation, the actual coordinates of the detector arrays are transferred to the CORSIKA coordinate system.
The program of photons ray-tracing is used for WFCTA simulation.
The trigger pattern of the telescope is set to 7, 
namely seven neighboring pixels should be triggered.
The fast simulation program~\cite{sim-wcda++} is used for the simulation of WCDA++ and MD.
Only photoelectrons are sampled according to the energy, direction and particle id of the secondary particles in EAS.
The time response is not included.
Then the simulated results of different detectors are merged before reconstruction.

\subsection{Event Reconstruction}

According to the actual data acquisition, as long as the Cherenkov telescope is triggered, the events measured by three detector arrays are reconstructed.

Firstly, the shower core position is given by WCDA++ through NKG fitting~\cite{nkg}.
Due to the lack of time response in WCDA++ fast simulation program, there is no information about the shower fronts.
Therefore the arrival direction of the shower is obtained by a Gauss sampling, 
of which angular resolution is 0.3$^{\circ}$.

Secondly, the perpendicular distance between the telescope and the shower axis ($R_p$) is calculated.
Next is Cherenkov image cleaning. 
The first step is to remove the fired pixels whose photoelectrons measured is less than 30 in the image.
The second step is to find the fired pixel with the largest number of photoelectrons,
and use it as the center to traverse the whole image outward. 
Then remove all the islanded fired pixels.

After image cleaning, the Hillas parameters~\cite{Hillas}, 
are given, as well as the total photoelectrons ($N^{pe}$) in the image.
$N^{pe}$ is a good energy estimator.
Before energy reconstruction, $N^{pe}$ should be normalized to $R_p=0$ and $\alpha=0$, 
namely
\begin{equation}
N_{0}^{pe}=lg(N^{pe})+a{\times}(R_p/1m)+b{\times}tan(\alpha)
\end{equation}
where $\alpha$ is the space angle between the shower direction and the optical axis of the telescope,
and the parameters $a$ and $b$ depend on the primary component of cosmic rays.

Finally, the lateral distribution of muons is fitted by an empirical formula Eq.(~\ref{muonlateral}) through maximum likelihood fitting~\cite{EAS}.
The normalized muon size is given.
\begin{equation}
{\rho}(r,N_{\mu}) = k_G N_{\mu} (\frac{r}{r_G})^{-k_1} (1+\frac{r}{r_G})^{-k_2}     [m^{-2}]
\label{muonlateral}
\end{equation}
where $k_1\cong1.4$, $k_2\cong1.0$, $r_G\cong220m$.
These three parameters are reconfirmed according to the layout of MD array. 

\subsection{Event Selection}

There are three basic principles for events selection:
firstly, the shower core position falls in WCDA++;
secondly, the Cherenkov image is complete;
thirdly, the muon lateral distribution is well fitted.

In detail, the reconstructed shower core position located in $130 m\times130 m$ arround WCDA++ is selected.
For a Cherenkov image, the number of fired pixels are more than 10;
and the angular distance between the weighted center of the image and the center of the camera plane is less than $6^{\circ}$.
For muon lateral distribution fitting, 
the fitted normalized muon size ($k_G N_{\mu}$) should be more than $10^{-7}$.

\begin{figure}[htb]
	\centering
	\includegraphics[width=0.49\linewidth]{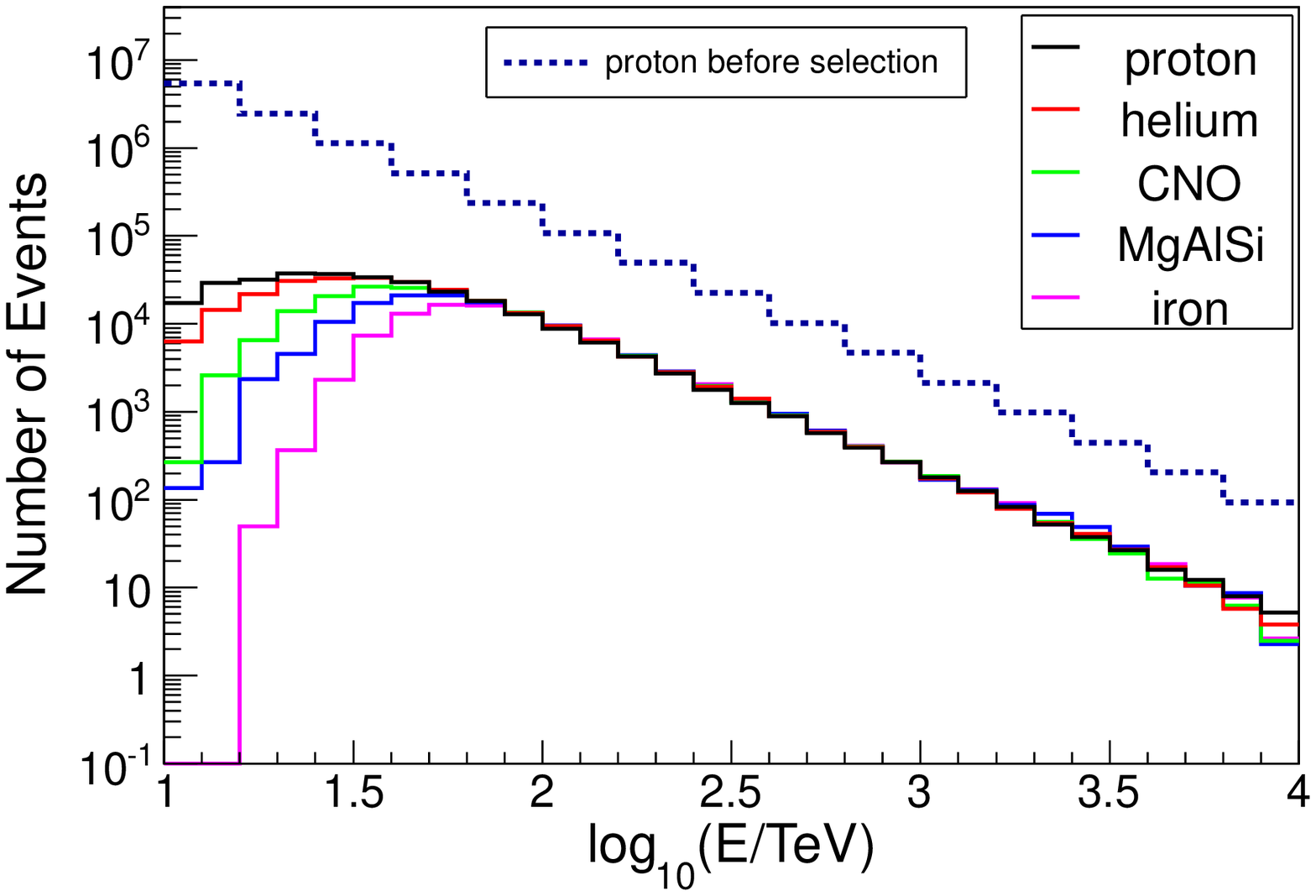}
	\includegraphics[width=0.49\linewidth]{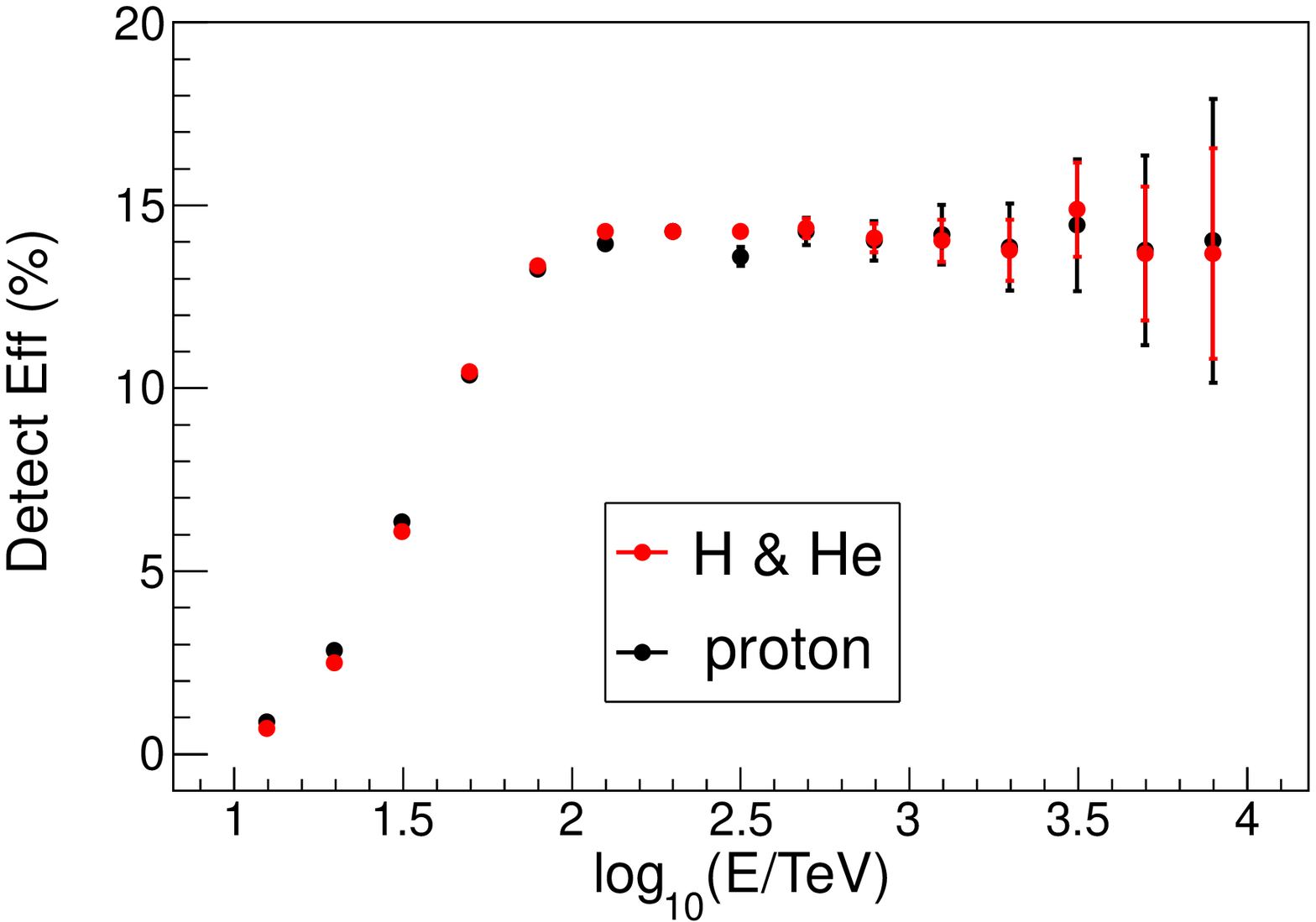}
	\caption{Left: the energy distribution after event selection.  Right: the detection efficiency of proton and H\&He.}
	\label{detection_eff}
\end{figure}

The number of events in each energy bin after selection is shown in Fig.~\ref{detection_eff} (left).
The dotted line represents the number of proton events involved in the simulation, 
which is consistent with the black line in Fig.~\ref{raw_specra}.
The solid lines represent the number of events after selection, 
and different colors represent different components.
It points out that the hybrid observation becomes
nearly fully efficient above 100 TeV for all components.

Detection efficiency is defined as the ratio of the number of selected events to the number of simulation events in each energy bin.
That is the black solid line divided by the blue dotted line in Fig.~\ref{detection_eff} (left).
In this simulation, the detection efficiency is about 15\% above 100 TeV for all components, 
as shown in Fig.~\ref{detection_eff}(right).
The red dots show the detection efficiency of H\&He and 
the black dots show the the detection efficiency of proton.
Besides, it is obvious that the error bars of the last two points 
in Fig.~\ref{detection_eff} (right) are large due to the small statistics.
Therefore the results of $lg(E/TeV) > 3.6$ are masked in the subsequent analysis.

\section{Proton and Helium Event Selection}\label{section:particle identification}

In this section, the parameters of the three types of detector arrays are analyzed according to the development characteristics of the EAS.
The component sensitive parameters of LHAASO hybrid observation are introduced.
Then, the Toolkit for Multivariate Analysis (TMVA)~\cite{tmva,tmva1} is used and the selection results are presented.

\subsection{Component Sensitive Parameters}

\subsubsection{Parameters from WFCTA}
It is known that iron-induced air showers develop larger and faster for the smaller interaction mean free path while primary particles travel into the atmosphere~\cite{EAS}.
Thus, the proton initiated showers develop to its maximum later than one of iron showers. 
The atmosphere depth of shower maximum, $X_{max}$, is a traditional mass sensitive parameter.
Here, ${\Delta}{\theta}$, the exactly parameter used to reconstruct the $X_{max}$, is applied instead of the reconstructed $X_{max}$.
${\Delta}{\theta}$ is the angular distance between the shower arriving direction and the gravity center of the Cherenkov image. 
But it can not be used directly to classify primary particles because of the $R_{p}$ and shower energy ($N_{0}^{pe}$) dependence.
After normalization, the structure of the mass sensitive parameter $P_{X}$ is as follows:
\begin{equation}
P_{X}={\Delta}{\theta}-0.0103\times R_p - 0.404\times N_{0}^{pe}
\end{equation}

Moreover, the proton-induced showers exhibit an elongated elliptic shape~\cite{Hillas}. 
And the ratio of length and width of the cherenkov image is also a traditional and effective parameter:
\begin{equation}
P_{C}=L/W-0.0137\times R_p+0.239\times N_{0}^{pe}
\end{equation}

\begin{figure}[htb]
	\centering
	\includegraphics[width=0.45\linewidth]{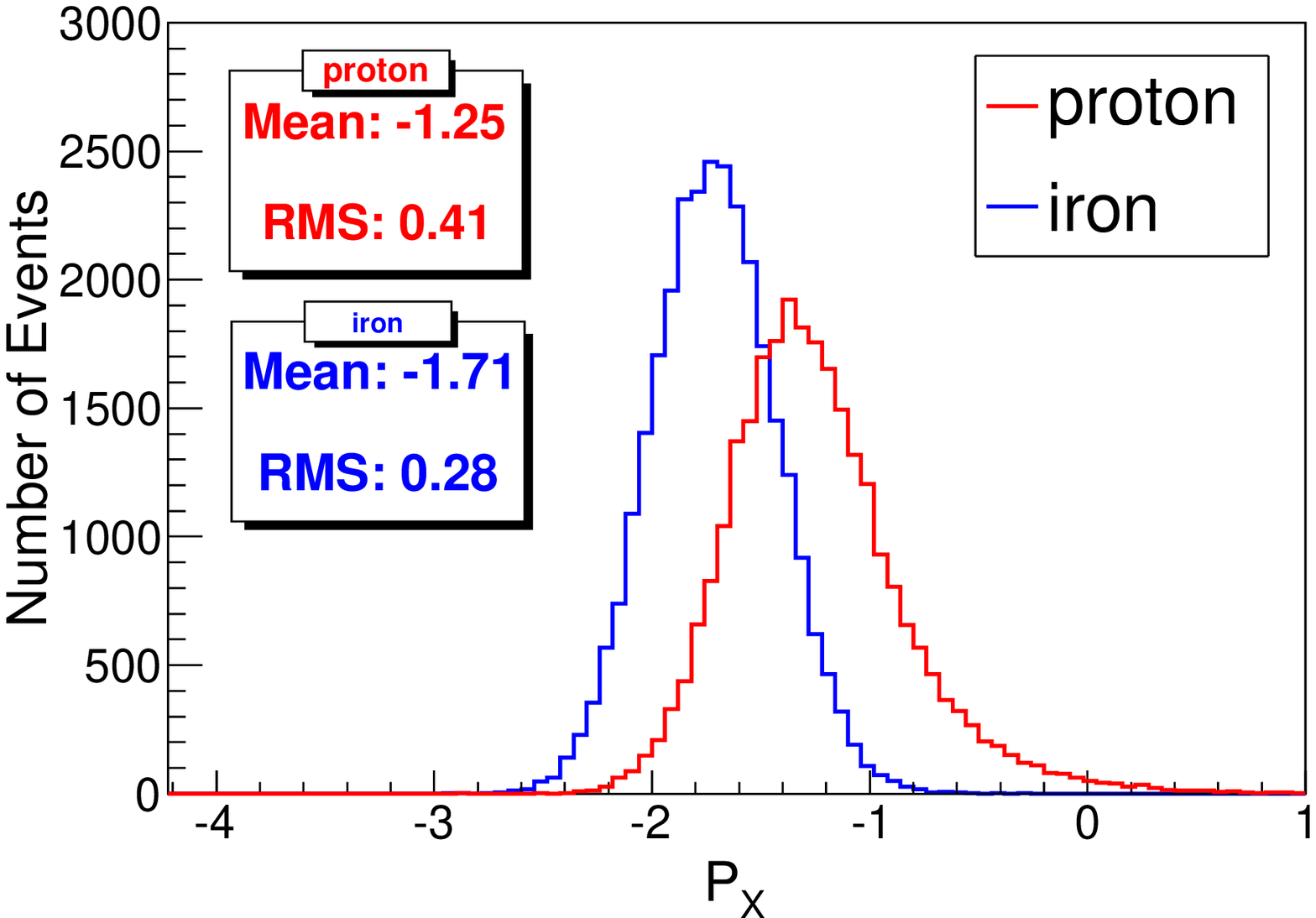}
	\includegraphics[width=0.45\linewidth]{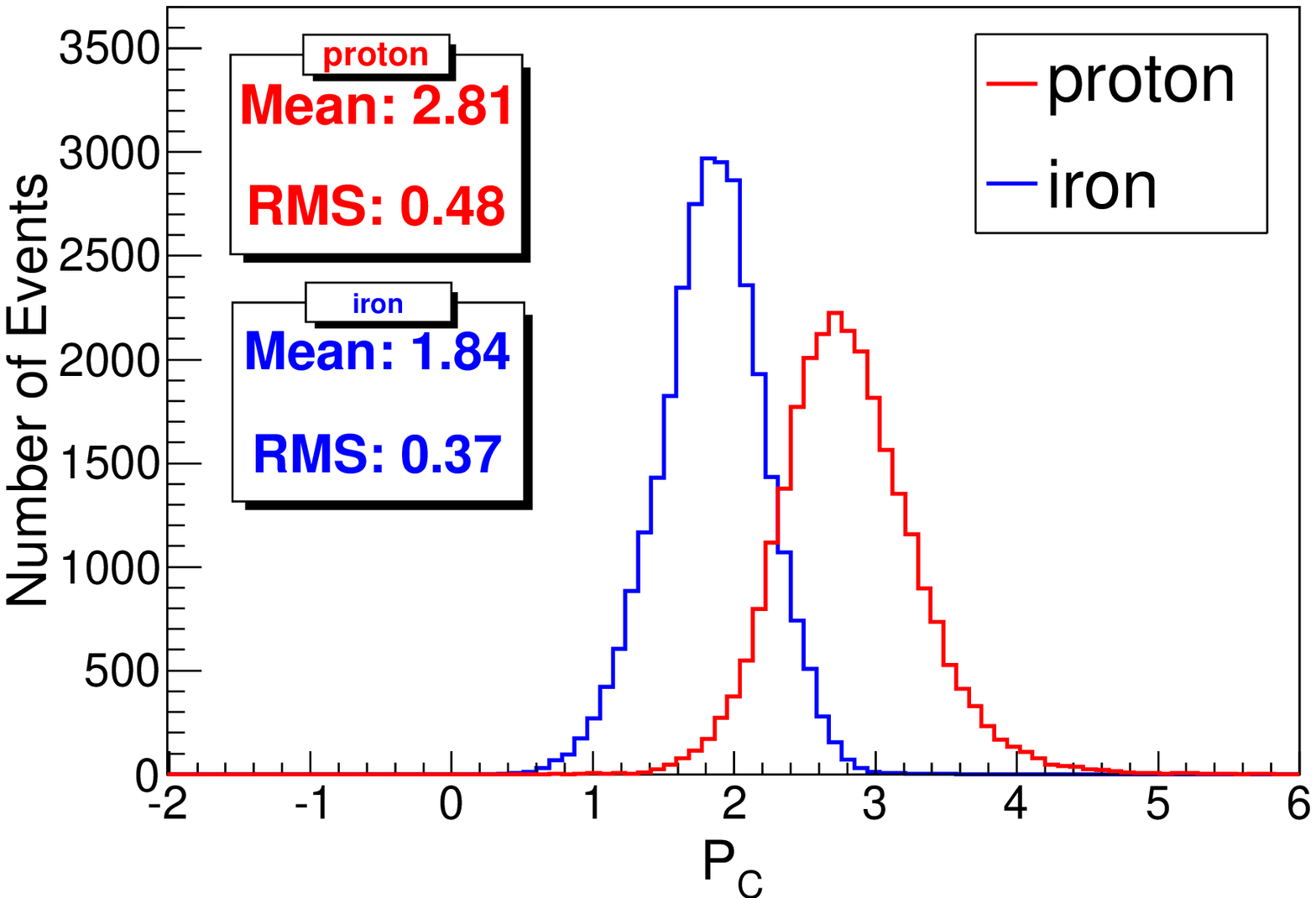}
	\caption{The distributions of mass sensitive parameters $P_{X}$ (left) and $P_{C}$ (right) for proton (red line) and iron (blue) initialed showers.}
	\label{pxpc}
\end{figure}

The distributions of $P_{X}$ and $P_{C}$ for proton and iron showers are shown in Fig.~\ref{pxpc}.

\subsubsection{Parameters from MD}
The muon size $N_{\mu}$ of a shower heavily depends on the atomic number ($A$) of the primary particle:
$N_{\mu}^{A}/N_{\mu}^{p} \approx A^{(1-\eta )}$, where $\eta$ is approximately 0.9~\cite{EAS} and $p$ indicates the proton shower.

Therefore, the muon size by fitting ($N_{\mu}^{F}$), the total number of detected muons ($N_{\mu}^{M}$) and the number of fired MDs (NMD) are significant variables to identify the primary particles:
\begin{equation}
P_{\mu1} = lg(N_{\mu}^{F})-0.9823 \times N_{0}^{pe}
\end{equation}
\begin{equation}
P_{\mu2} = lg(N_{\mu}^{M})+0.00226 \times Rlp-0.873 \times N_{0}^{pe}
\end{equation}
\begin{equation}
P_{\mu3} = lg(NMD+0.098\times Rlp)-0.552\times N_{0}^{pe}
\end{equation}
$Rlp$ is the perpendicular distance between the center of the MD array and the shower axis.
The distributions of three parameters $P_{\mu1}$, $P_{\mu2}$ and $P_{\mu3}$ for proton and iron showers are shown in Fig.~\ref{pfpm}.

\subsubsection{Parameters from WCDA}

It is also known that the iron initialed shower is more extensive in the same observation level. 
It means that the lateral extension of secondary particles in iron-induced air shower is larger. 
WCDA++, the fully covered detector array, can well detect the lateral distribution of secondary particles.
The formulas for the average lateral distribution $<ER>$ and fluctuation $RMS$ are expressed as Eq. 2 to Eq. 5.

\begin{equation}
<ER> = \dfrac{\sum R_{i} \times Pe_{i}}{\sum Pe_{i}} 
\label{ER}
\end{equation}
where $R_{i}$ is the distance between the shower core position and the $ith$ fired cell in WCDA++.
$Pe_{i}$ is photoelectronic measured by the $ith$ fired cell.
\begin{align}
	RMS_{x} = \sqrt{ \dfrac{\sum x_{i}^{2}\times Pe_{i}}{\sum Pe_{i}} - x_{weight}^{2} } \\
	RMS_{y} = \sqrt{ \dfrac{\sum y_{i}^{2}\times Pe_{i}}{\sum Pe_{i}} - y_{weight}^{2} } \\
	RMS = \sqrt{ \dfrac{\sum R_{i}^{2}\times Pe_{i}}{\sum Pe_{i}} - <ER>^{2} } 
	\label{rms}
\end{align}
where the $x_{weight}$ and $y_{weight}$ are the centroid of distribution of secondary particles in WCDA++: 
\begin{align}
	x_{weight} = \dfrac{\sum x_{i}\times Pe_{i}}{\sum Pe_{i}}\\
	y_{weight} = \dfrac{\sum y_{i}\times Pe_{i}}{\sum Pe_{i}}
	\label{x-y-weight}
\end{align}
These parameters are not only related to the primary component but also to the shower direction and the shower energy, 
hence it should be corrected before MVA: 
\begin{equation}
P_{F2} = lg<ER>-0.343\times \theta_{rec}+0.1159\times N_{0}^{pe}
\end{equation}
\begin{equation}
P_{F3} = RMS_{x}-6.921\times \theta_{rec}+2.82\times N_{0}^{pe}
\end{equation}
\begin{equation}
P_{F4} = RMS-6.084\times \theta_{rec}+1.917\times N_{0}^{pe}
\end{equation}

\begin{figure}[]
	\centering
	\includegraphics[width=0.49\linewidth]{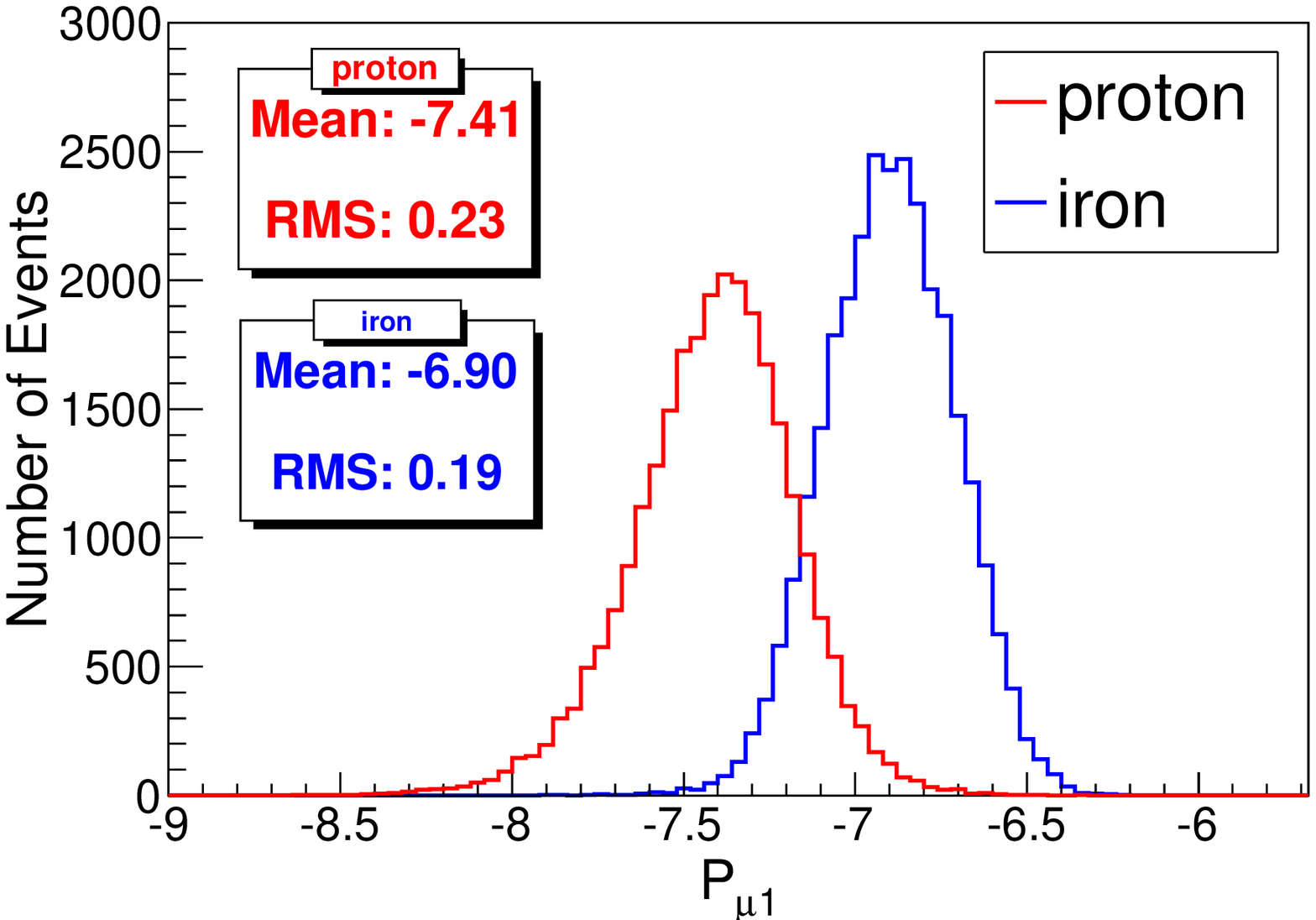}
	\includegraphics[width=0.49\linewidth]{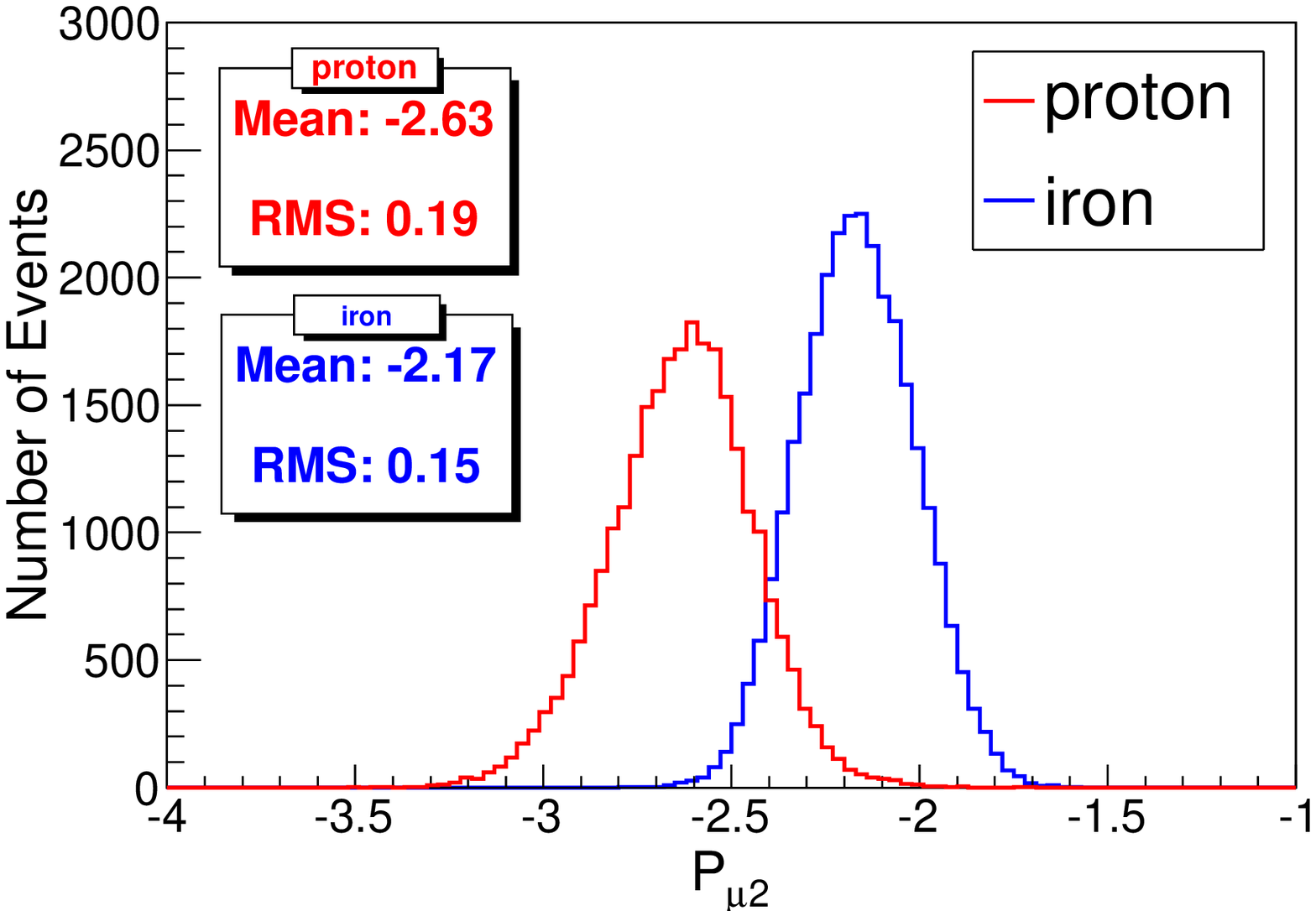}
	\includegraphics[width=0.49\linewidth]{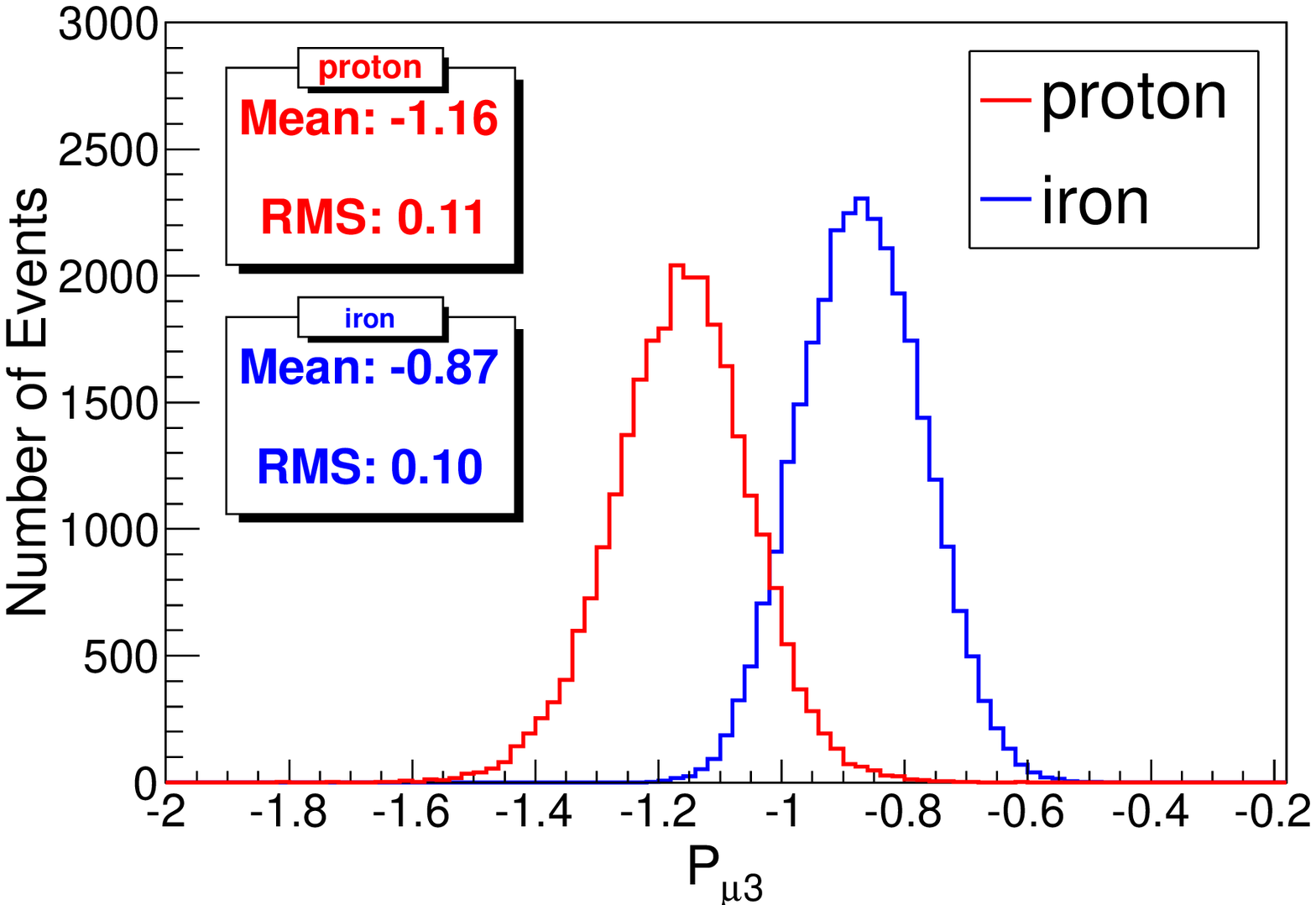}
	\includegraphics[width=0.49\linewidth]{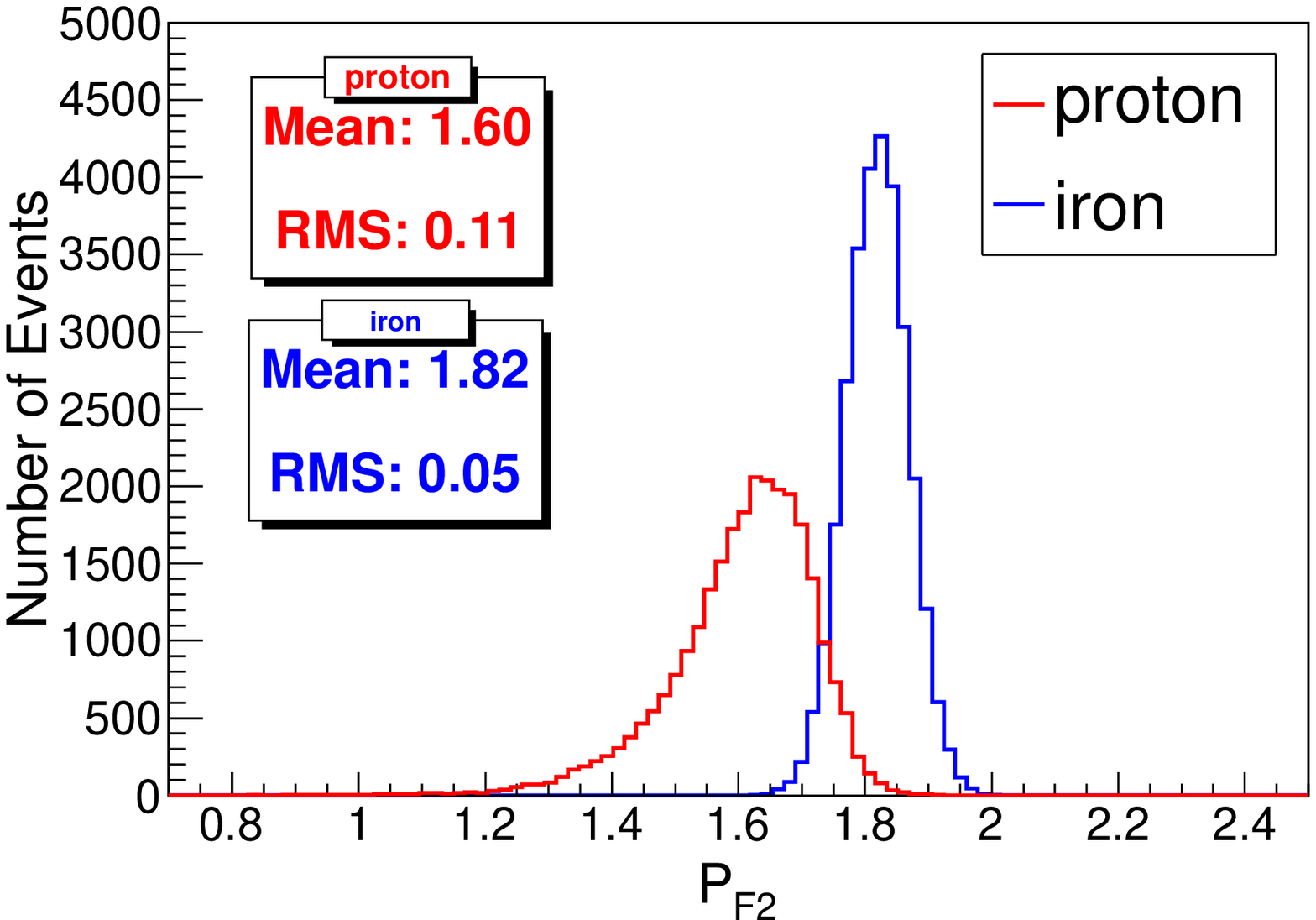}
	\includegraphics[width=0.49\linewidth]{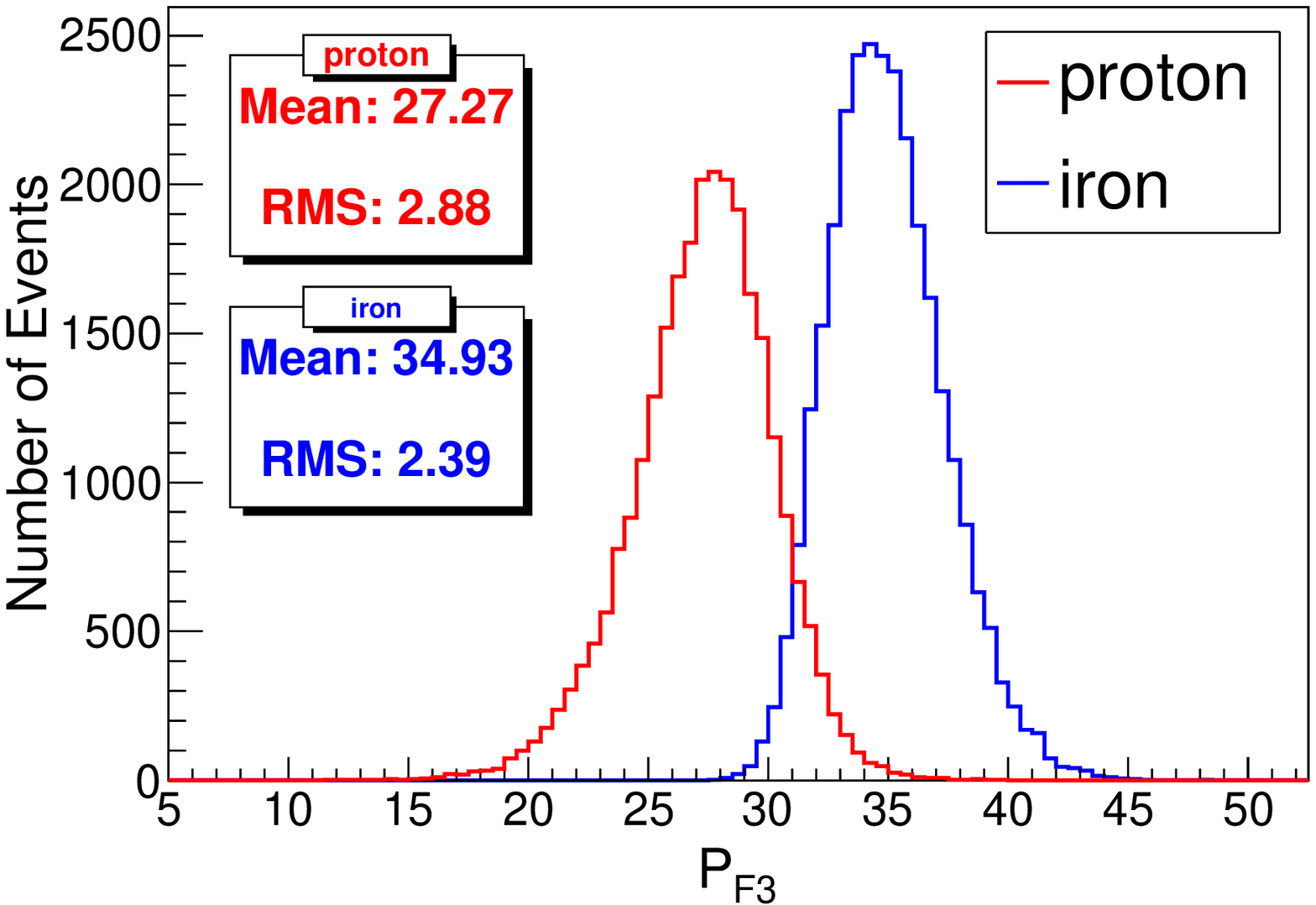}
	\includegraphics[width=0.49\linewidth]{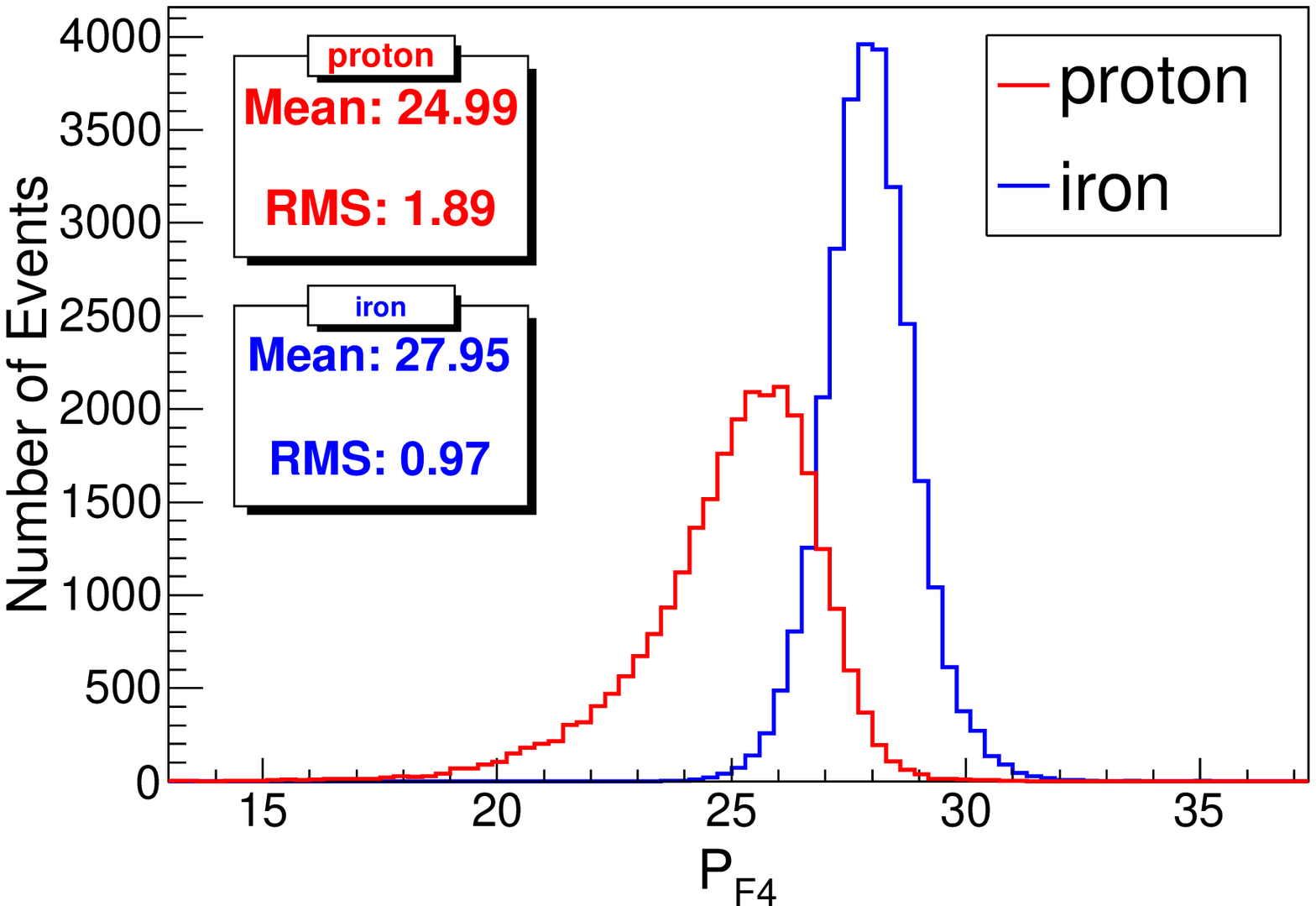}
	\caption{The distributions of mass sensitive parameters $P_{\mu1}$ (up-left), $P_{\mu2}$ (up-right), $P_{\mu3}$ (middle-left), $P_{F2}$ (middle-right), $P_{F3}$ (bottom-left) and $P_{F4}$ (bottom-right) for proton (red line) and iron (blue) initialed showers.}
	\label{pfpm}
\end{figure}

The distributions of three parameters $P_{F2}$, $P_{F3}$ and $P_{F4}$ for proton and iron showers are shown in Fig.~\ref{pfpm}.

\begin{figure}[htb]
	\centering
	\includegraphics[width=0.49\linewidth]{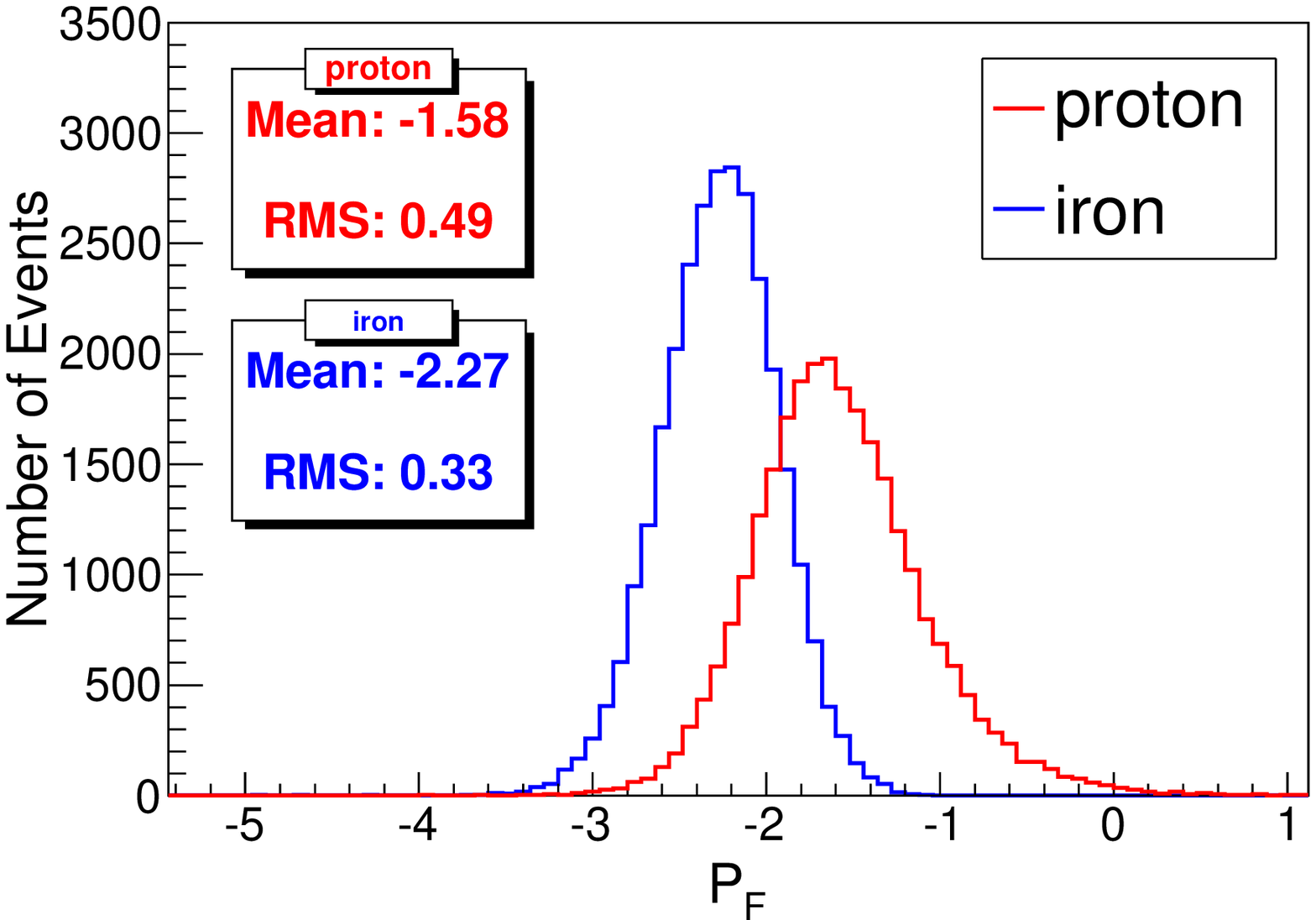}
	\includegraphics[width=0.49\linewidth]{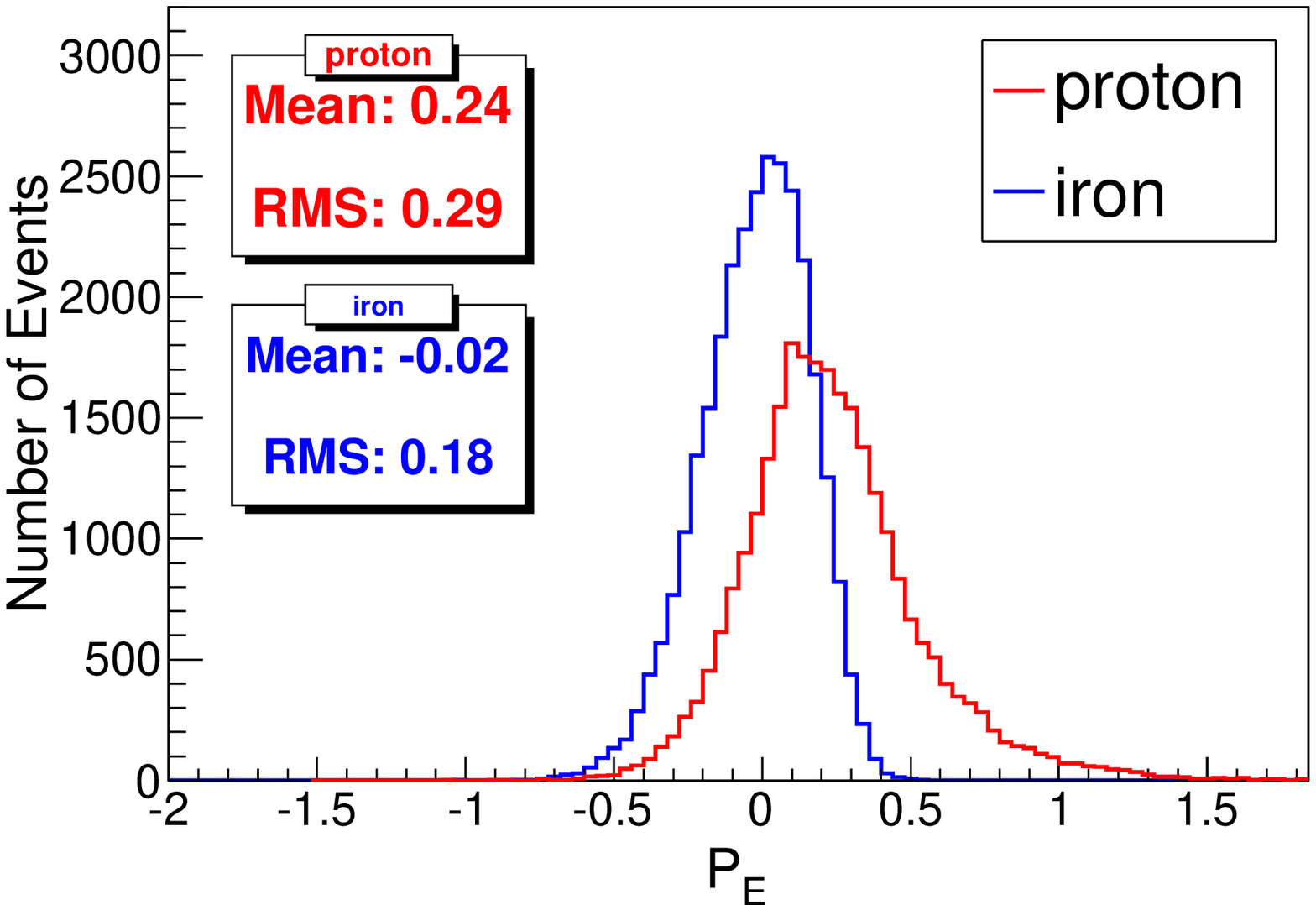}
	\caption{The distributions of mass sensitive parameters $P_{F}$ (left) and $P_{E}$ (right) for proton (red line) and iron (blue) initialed showers.}
	\label{pf}
\end{figure}

In addition, nearby the shower core axis,  the particle density of iron-shower is much less than that of proton-shower~\cite{argo-wfcta}.
Hence the photoelectrons in the brightest cell ($N_{max}$) measured by WCDA++ is sensitive to components: 
\begin{equation}
P_{F} = lg(N_{max}) -1.391\times N_{0}^{pe}
\end{equation}
The total photoelectrons ($N_{w}^{pe}$) measured by WCDA++ is also a mass sensitive variable, even it is weaker than $N_{max}$: 
\begin{equation}
P_{E} = lg(N_{w}^{pe}) -1.163\times N_{0}^{pe}
\end{equation}
The distributions of two parameters $P_{F}$ and $P_{E}$ for proton and iron showers are shown in Fig.~\ref{pf}.

Other parameters have also been studied, 
such as the photoelectrons measured by the brightest pixel in air Cherenkov image ($P_{C1}$), 
the total photoelectrons located 45 meters away from the shower core in the WCDA++ ($P_{F1}$), 
the muon density detected from 80 m to 100 m away from the shower core position ($P_{\mu4}$), and etc.
Nevertheless, the particle classification of these variables is weak.
After tuning of parameters, these ten parameters are candidates used as input parameters for the BDTG classifiers.

\subsubsection{Correlation of Parameters}
\begin{figure}[htb]
	\centering
	\includegraphics[width=0.45\linewidth]{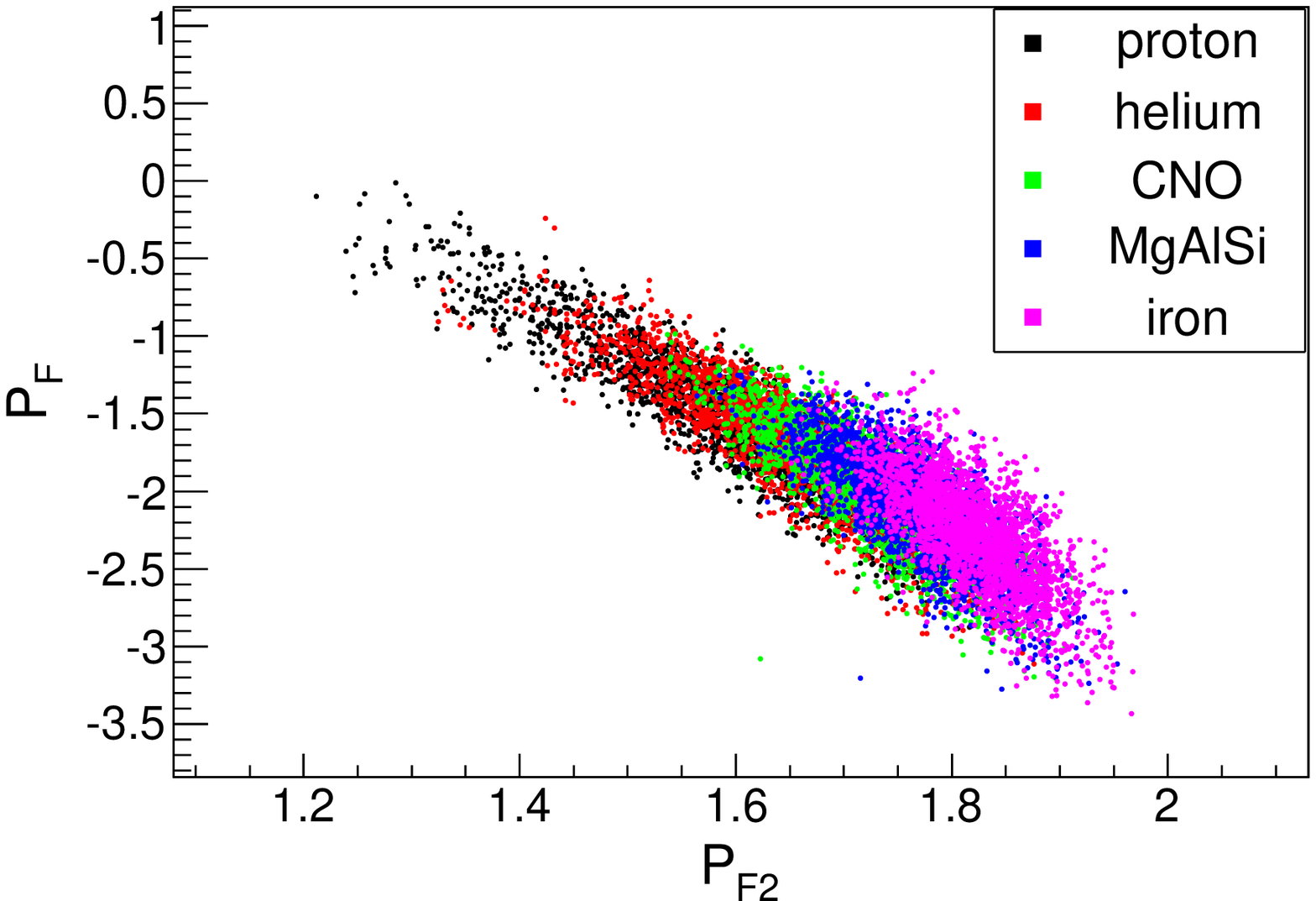}
	\includegraphics[width=0.45\linewidth]{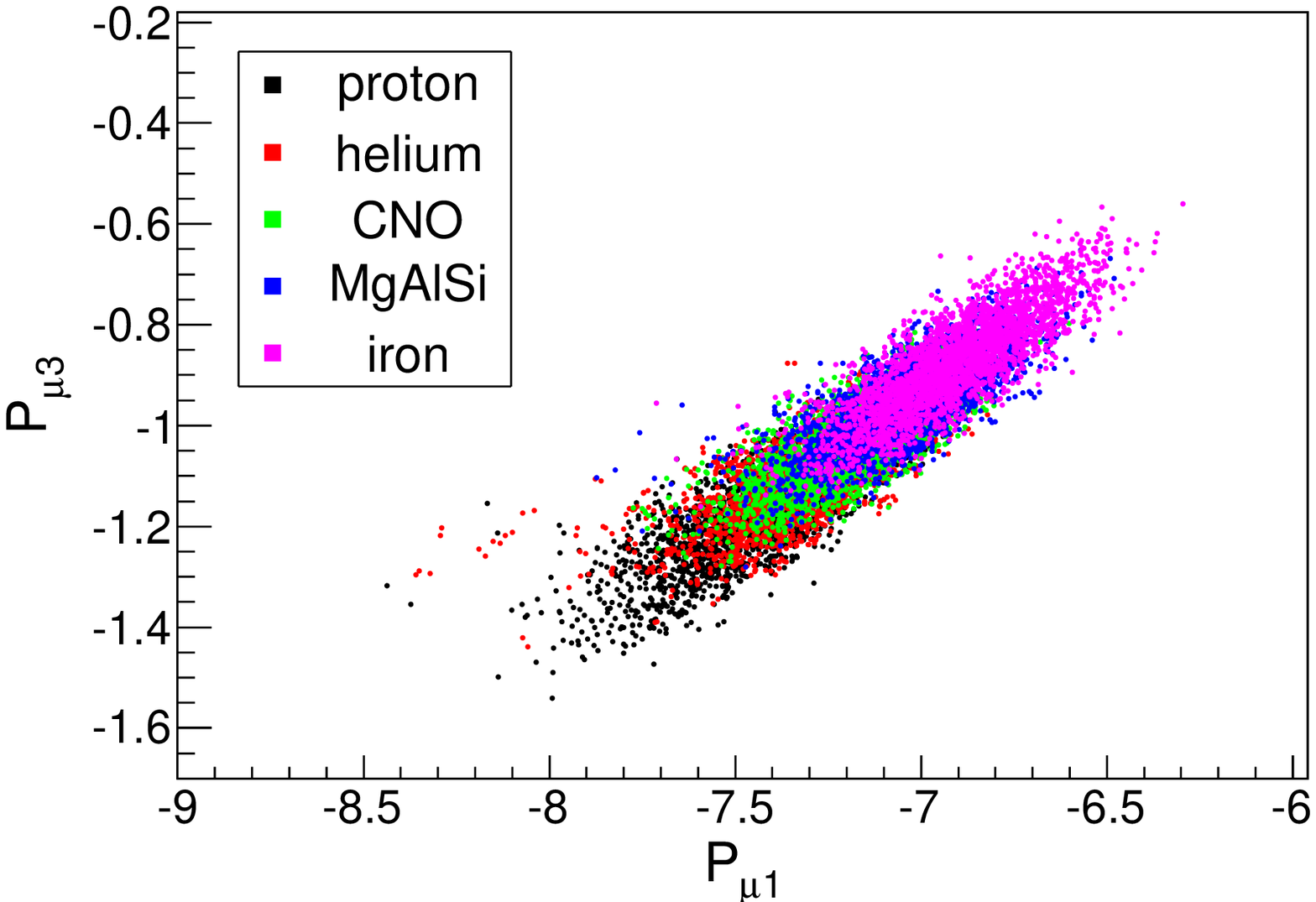}
	\caption{The correlations of mass sensitive parameters from WCDA (left) and MD (right).}
	\label{pfpmu}
\end{figure}

The parameters described above are not independent. 
There is a correlation between some parameters.
Parameters provided by the same detector array are highly correlated, such as, $P_{F}$ and $P_{F2}$, $P_{\mu1}$ and $P_{\mu3}$.
The correlations of parameters from WCDA and MD for five components of cosmic rays are shown in Fig.~\ref{pfpmu}.
The parameters of $P_{F}$ and $P_{F2}$ are negative correlation (left plot) and  $P_{\mu1}$ and $P_{\mu3}$ are position correlation (right plot).
According to the development characteristics of the EAS, it is easy to understand that the more extensive of the lateral distribution of the secondary particles, the less particles nearby the shower core axis; and the more muon detectors fired, the more muon contents in the shower.

\begin{figure}[htb]
	\centering
	\includegraphics[width=0.45\linewidth]{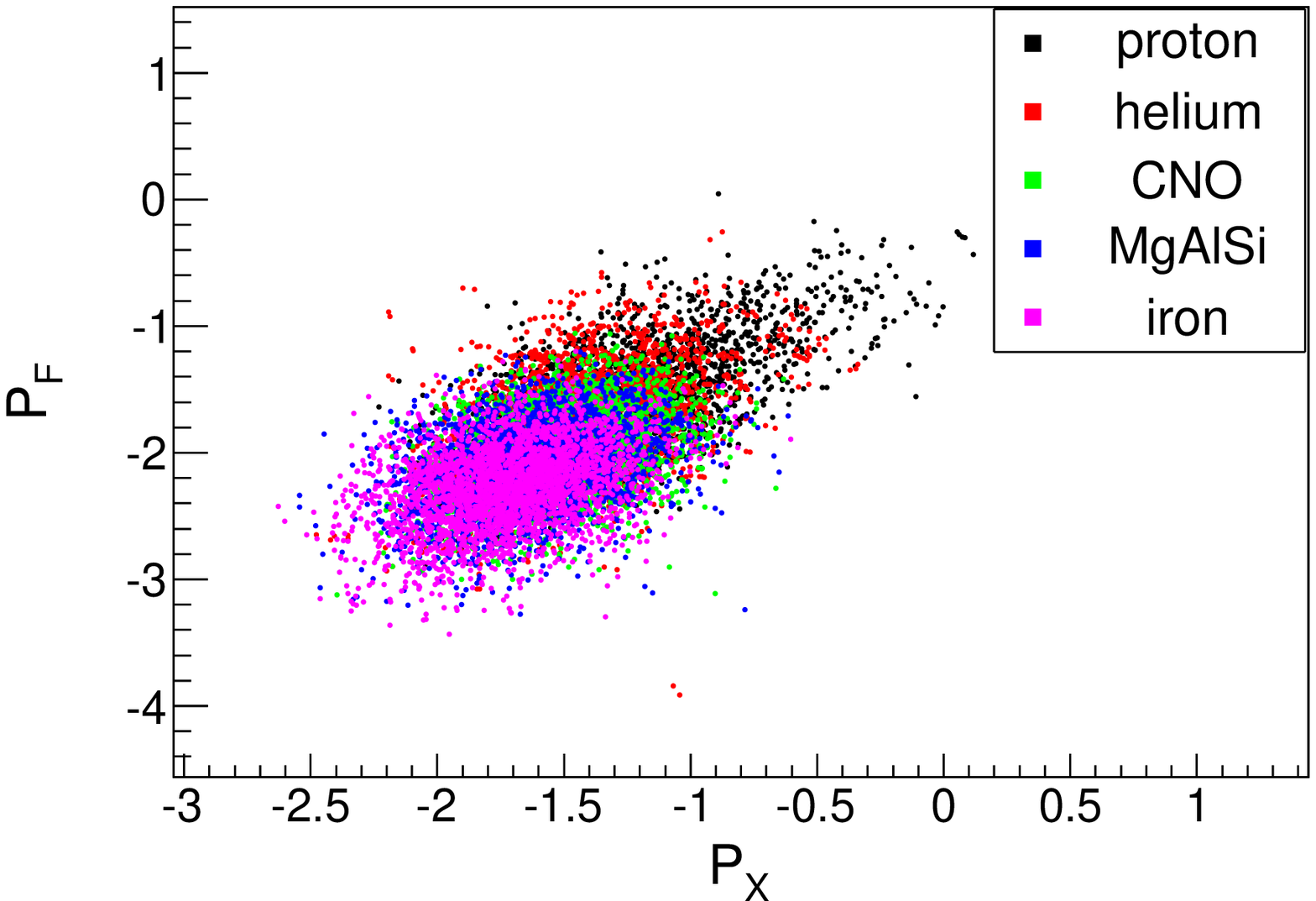}
	\includegraphics[width=0.45\linewidth]{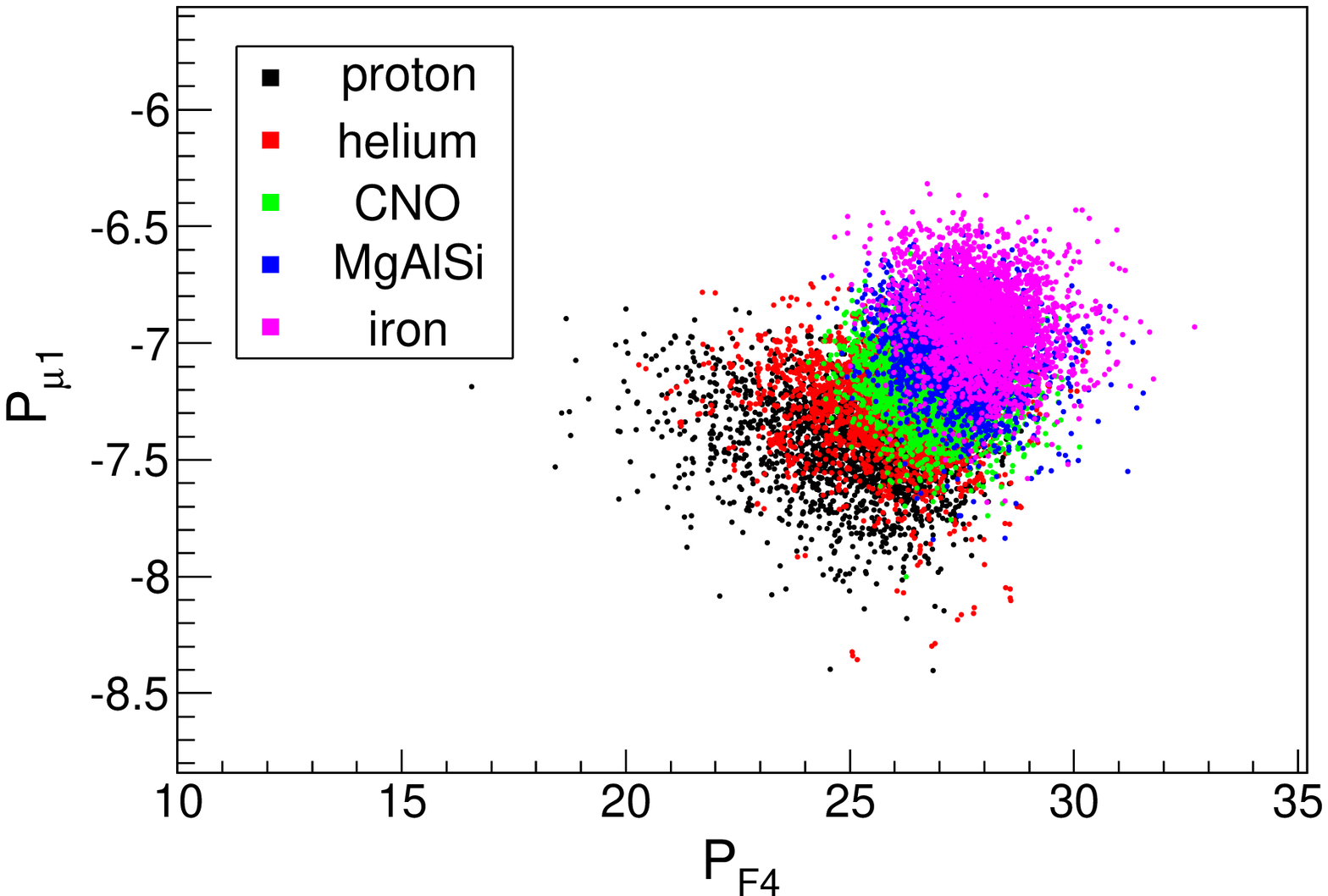}
	\caption{The correlations of mass sensitive parameters from different detector arrays.}
	\label{pxpfpm}
\end{figure}

The correlation of parameters provided by the different detector arrays is weak, such as, $P_{F}$ and $P_{X}$, $P_{\mu1}$ and $P_{F4}$.
The correlations of parameters from different detector arrays for five components are shown in Fig.~\ref{pxpfpm}.
The distributions are approximate to circle.
However, the correlation between parameters of different components is slightly different;
parameters for proton tend to be more correlated, as shown by the black dots in Fig.~\ref{pxpfpm}.

\begin{figure}[htb]
	\centering
	\includegraphics[width=0.5\linewidth]{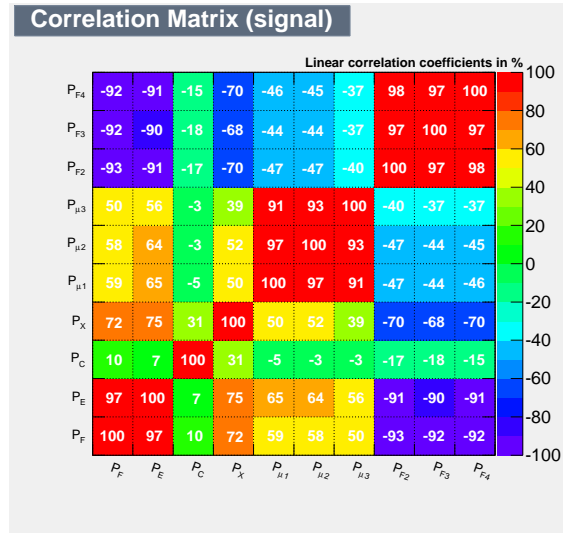}
	\caption{Linear correlation matrix for the input variables of proton events.}
	\label{correlation}
\end{figure}

The correlation matrix of these ten variables for proton initialed showers is shown in Fig.~\ref{correlation}.
The linear correlation coefficient 100\% means a linear positive correlation and -100\% means linear negative correlation. 0 means no correlations among two parameters.
It should be noticed that the parameter $P_{C}$ is basically independent of other parameters, as shown in the third column or in the eighth row in Fig.~\ref{correlation}.
This is because $P_{C}$ reflects the overall development characteristic, not just the lateral or longitude distribution of the air shower.

After removing the highly correlated (97\% or 98\% correlations) parameters, 
six parameters, $P_{C}$, $P_{X}$, $P_{F}$, $P_{F4}$, $P_{\mu2}$ and $P_{\mu3}$, 
are used for TMVA training and analysis.

\subsection{TMVA analysis}

Drawing on the experience of ARGO-YBJ \& WFCT hybrid detection, two parameters are used for particle identification~\cite{classification-old} in the event-by-event cut.
However, too many variables are no longer suitable for particle identification through event-by-event cut 
because it is complicated to get the optimal cut value of each parameter.
Therefore, MVA is taken.

As an important branch of statistics, multivariate analysis has been applied to most of the disciplines.
TMVA is specially developed for high energy physics based on the ROOT-integrated environment. 
It is powerful for signal and background classification.
In accelerator physics, it can effectively screen out the b-tagging signals from a large number of background particles in the jet~\cite{tmva-LHC}.
And likewise, it enables to identify the components of cosmic rays~\cite{tmva-zzz}.

TMVA works based on machine learning.
It integrates multiple advanced algorithm classifiers, such as Boosted Decision Trees (BDT), Artificial Neural Networks (ANN), Support Vector Machine (SVM), and etc. 
Users need to input variable samples and select the algorithm. 
Then the machine training and testing are carried out. 
The result is to output one variable for users to select signals.

Here Boosted Decision Trees with Gradient boosting(BDTG) is chosen, which is the most widely used so far.
The classification of $H$ and $H\&He$ from other mass components is carried out independently.

The selected hybrid events are divided into two equal parts.
One is used for TMVA training and testing;
the other is used as data.
The statistics of signal (proton or H\&He) and background (others) are shown in the Table ~\ref{no-train}.

\begin{table}[ht]
	\centering
	\caption{The number of events for BDTG classifier training and testing.}
	\begin{tabular}{|c|c|c|}
		\hline
		NO. of events & Proton & H\&He \\
		\hline
		Signal & 40215 & 61098\\
		\hline
		Background & 91330 & 70447\\
		\hline
	\end{tabular}
	\label{no-train}
\end{table}

To avoid overtraining, several parameters are adjusted to achieve the best performance of the classifier.
Parameters of BDTG classifiers for the separation of H/other nuclei are as follows: 
\begin{itemize}
	\item[-] Number of trees in the forest is 380;
	\item[-] Minimum training events required in a leaf node are 30;
	\item[-] Max depth of the decision tree is 2.
\end{itemize}
Parameters of BDTG classifiers for the separation of H\&He/other nuclei are as follow: 
\begin{itemize}
	\item[-] Number of trees in the forest is 300;
	\item[-] Minimum training events required in a leaf node are 50;
	\item[-] Max depth of the decision tree is 2.
\end{itemize}
The other parameters are set as default values.

\begin{figure}[htb]
	\centering
	\includegraphics[width=0.45\linewidth]{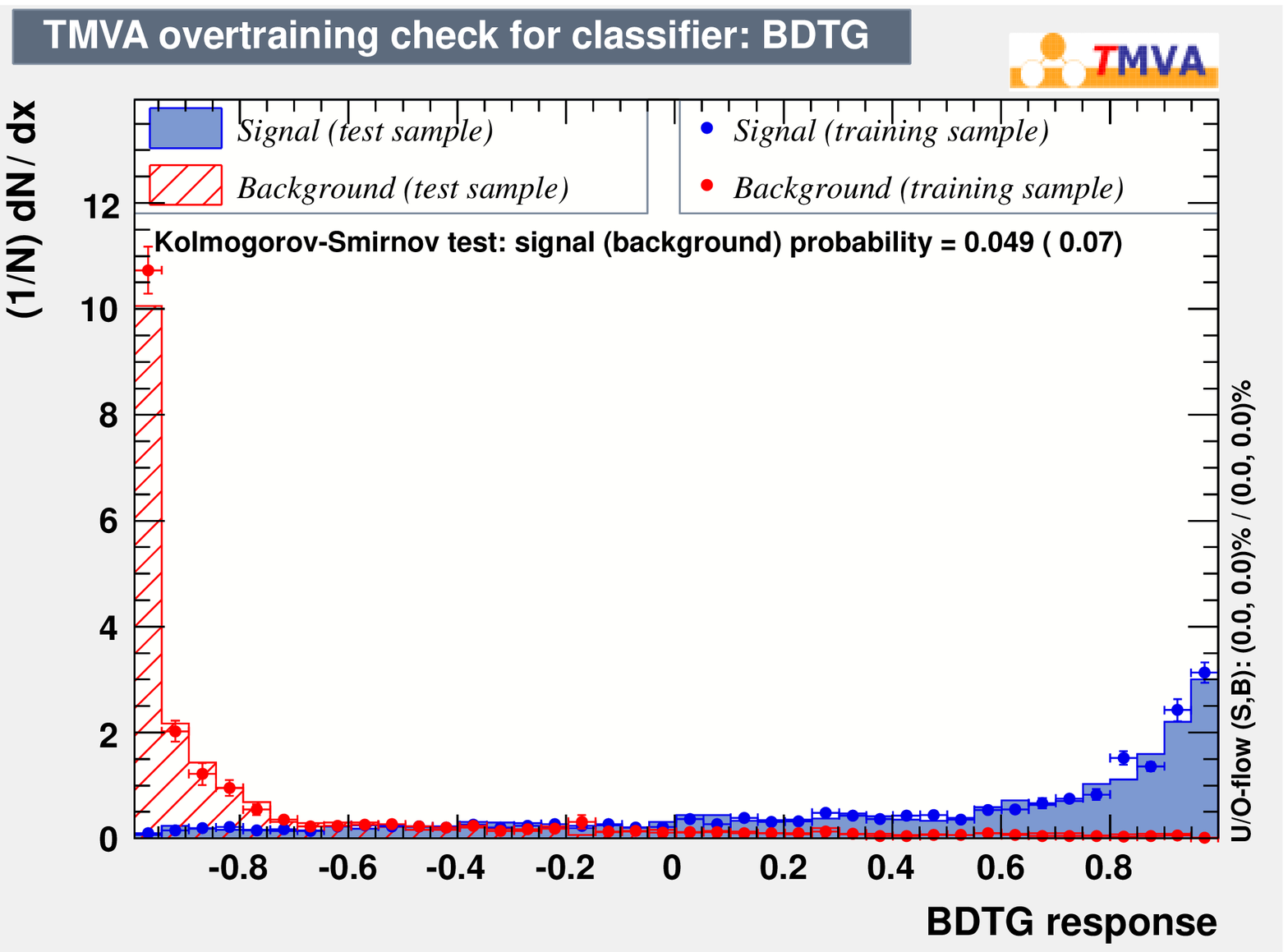}
	\includegraphics[width=0.45\linewidth]{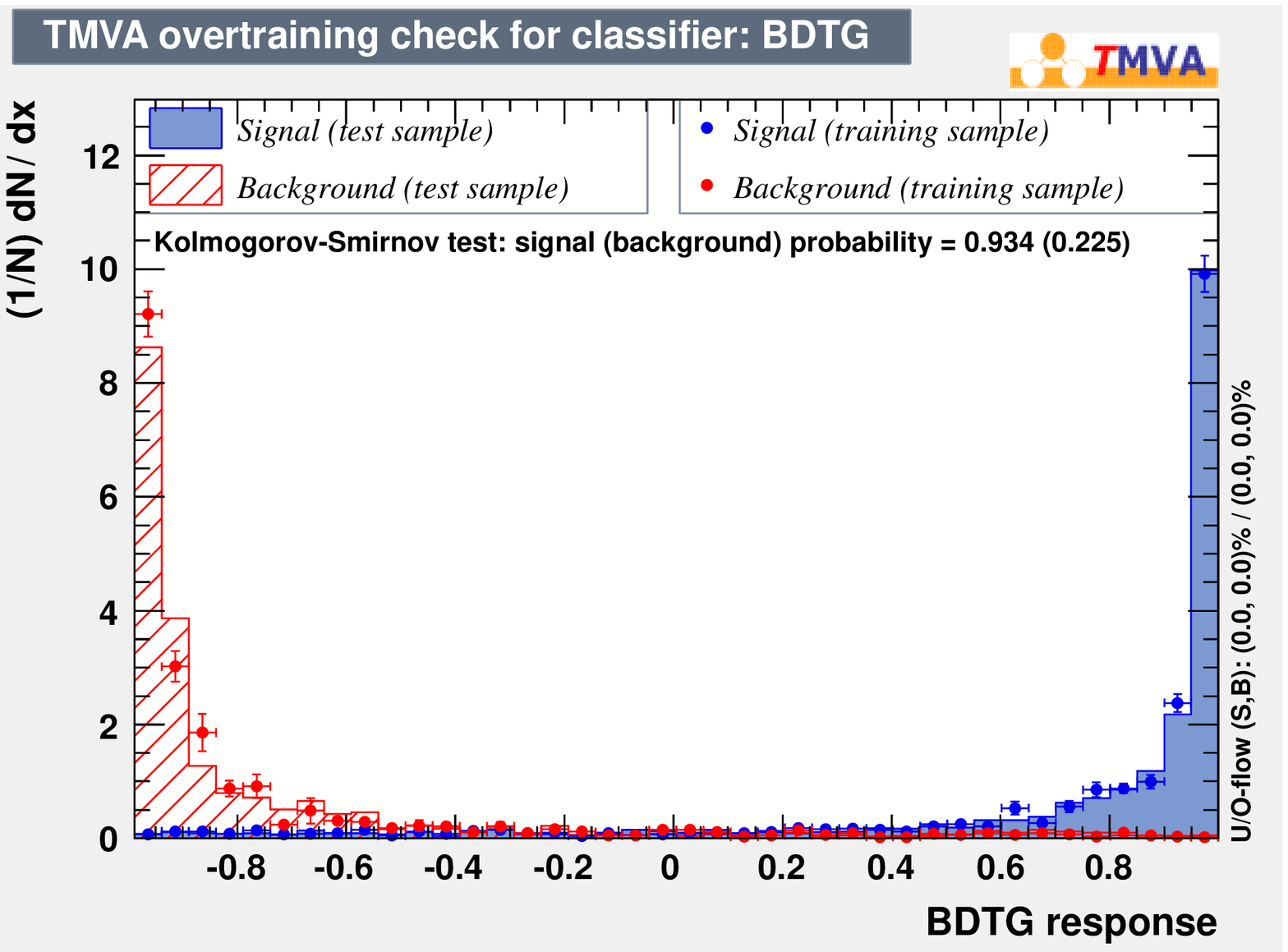}
	\caption{Training results of BDTG classifier for proton (left) and H\&He (right).}
	\label{trian-result}
\end{figure}

\begin{figure}[htb]
	\centering
	\includegraphics[width=0.5\linewidth]{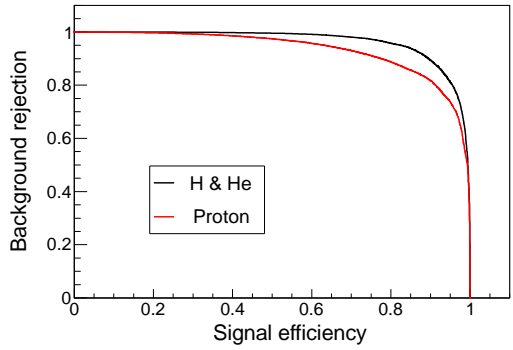}
	\caption{ROC curves BDTG classifier for proton (red line) and H\&He (black line).}
	\label{trian-roc}
\end{figure}

The training results are shown in Fig.~\ref{trian-result}.
$H\&He$ can be well separated from other components;
However, separation of proton from other nuclei is barely satisfactory.
The background rejection versus signal efficiency (``ROC curve") is obtained,
by cutting on the BDTG classifier outputs for the events of the proton and H\&He samples,
as shown in Fig.~\ref{trian-roc}.
It is obviously that under the same signal efficiency, 
the background rejection of other heavy nuclei (H\&He vs. other nuclei) is higher.

\subsection{Results}

\begin{figure}[htb]
	\centering
	\includegraphics[width=0.5\linewidth]{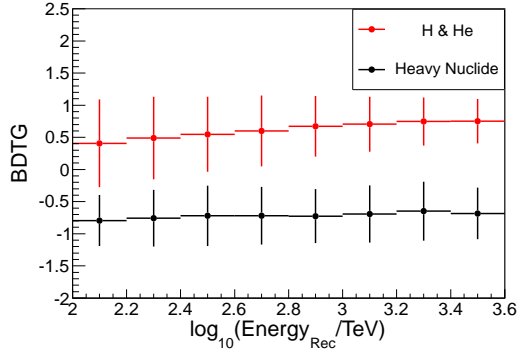}
	\caption{The distribution of output parameter BDTG as a function of reconstructed energy. The error bars in the figure shows the RMS in each bin.}
	\label{tBDTG-Erec}
\end{figure}

The training results are applied to the other half of the data sample.
Generally, the signal events can be selected according to the best cut points provided by TMVA.
However, because our final goal is to obtain the selection efficiency under different energy bins, 
the distribution of the output of BDTG classifier versus reconstructed energy is studied firstly.
As shown in Fig.~\ref{tBDTG-Erec}, the separation of the output BDTG between H\&He and heavy nuclei become larger as primary energy increases.
The mean value of the signal (red dots) in each energy bin is slightly away from the background (black dots) and the RMS of the signal is gradually decreased. 
The causes for the separation becoming larger is that the performance of the input parameters gets better in higher energies.

To get a coincident selection efficiency, the cut values are scaled following the Eq.~\ref{Hcut} and Eq.~\ref{HHecut}:
\begin{equation} 
\centering
Cut Value (H) = 0.09\times lg(Energy/TeV)+0.55
\label{Hcut}
\end{equation}
\begin{equation} 
\centering
Cut Value (H\&He) = 0.33\times lg(Energy/TeV)-0.26
\label{HHecut}
\end{equation}

\begin{figure}[htb]
	\centering
	\includegraphics[width=0.49\linewidth]{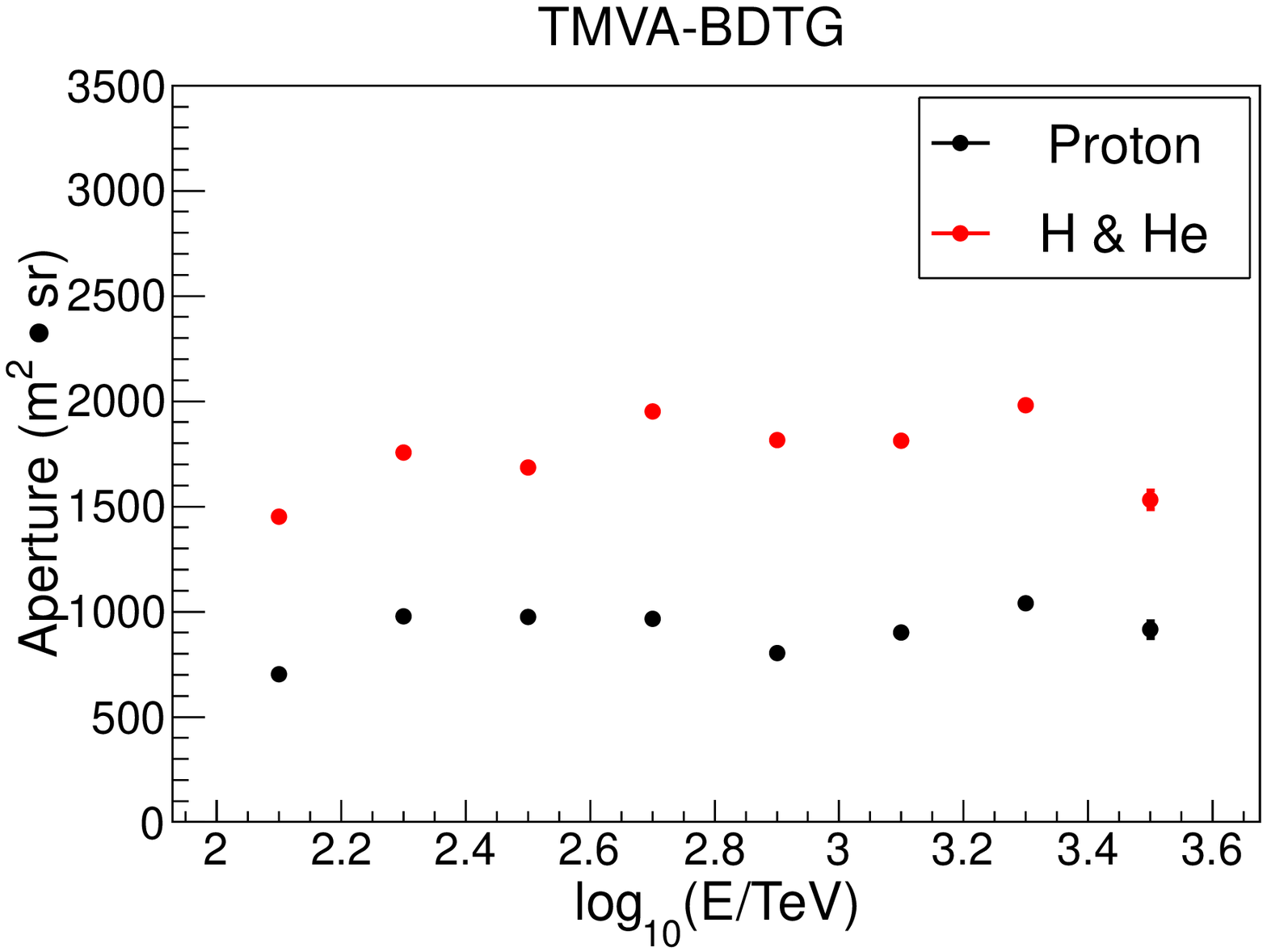}
	\includegraphics[width=0.49\linewidth]{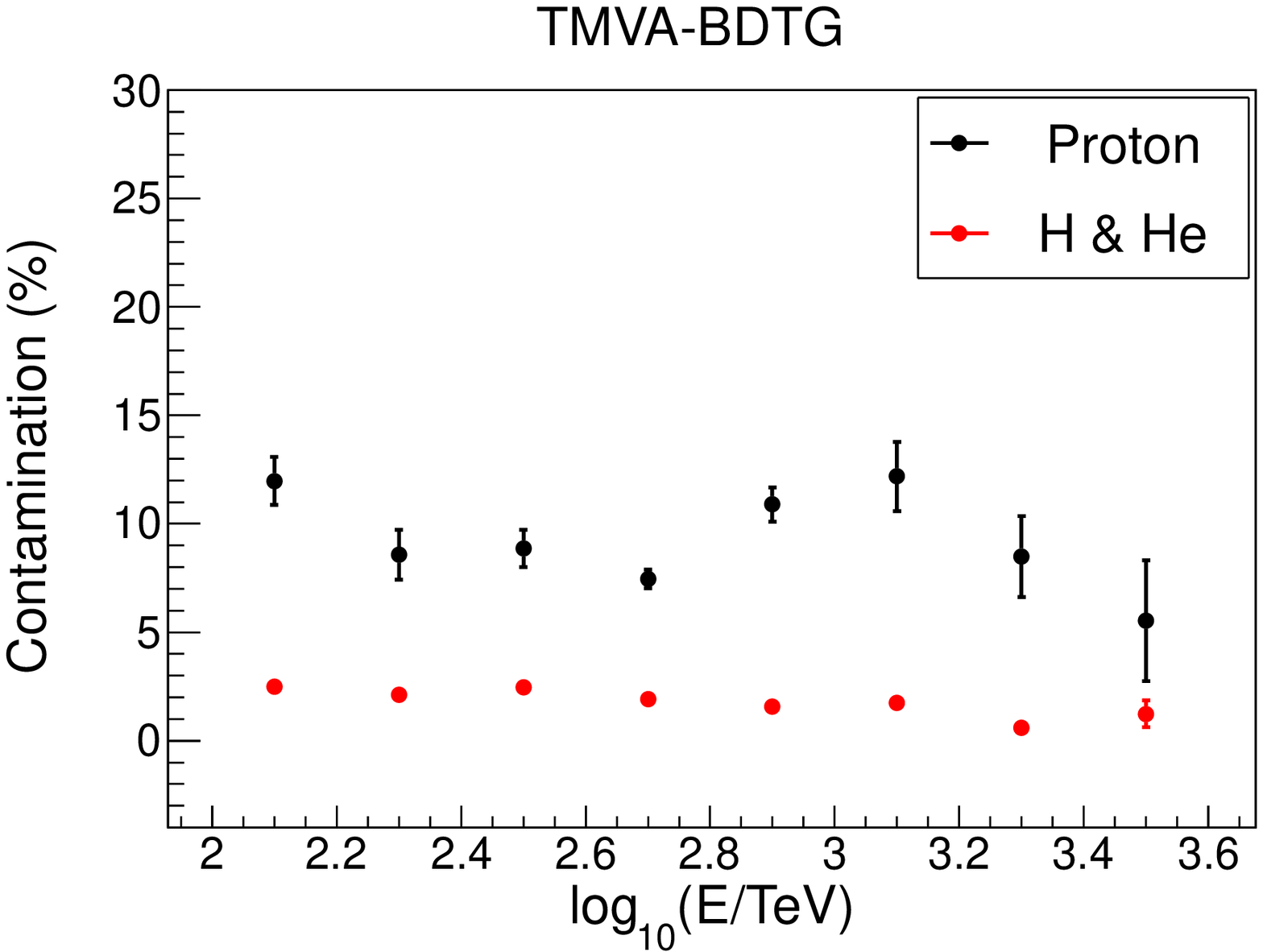}
	\caption{The final selected aperture (left) and contamination (right) for H and H\&He.}
	\label{apert}
\end{figure}

After the cut described above, the aperture and contamination of hybrid observation are calculated.
The aperture of pure proton is about $900 m^{2}Sr$  and the aperture of H\&He is about $1800 m^{2}Sr$,
as shown in Fig.~\ref{apert} (left).
The contamination of proton sample is about 10\% and the contamination of H\&He sample is less than 3\%
according to the H\"{o}randel model,
as shown in Fig.~\ref{apert} (right).

Considering the uncertainty due to the hadronic interaction models,
there are two batches of data, QGSJET-FLUKA and EPOS-FLUKA, were used for error analysis.
It is found that the aperture of H\&He is consistent within $\pm 5\%$.


\section{Proton and Helium Spectrum Expectation of LHAASO}\label{section:spectrum}

\begin{figure}[htb]
	\centering
	\includegraphics[width=0.5\linewidth]{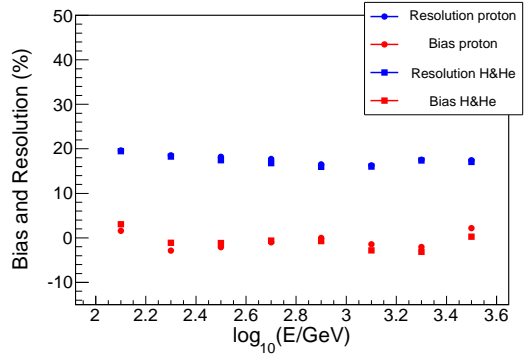}
	\caption{Bias and resolution of reconstructed energy for proton and H\&He.}
	\label{Proton_Rec}
\end{figure}

After the primary particle identification, the accuracy of event reconstruction for light components is obtained.
For proton and H\&He, the shower core resolution is less than 3 m.
The energy reconstruction is obtained by WFCTA, as described in Section 3.
There is a linear correlation between shower energy E and the corrected number of Cherenkov photoelectrons, $N_{0}^{pe}$.
For proton, $a=0.00916$ and $b=0.0182$;
the reconstructed energy is $E_{rec}=10^{0.9558{\times}N_{0}^{pe}-2.31}$.
The variables $a$ and $b$ for H\&He energy reconstruction are approximately equal to that of proton.
The reconstructed bias and resolution for proton and H\&He are shown in Fig.~\ref{Proton_Rec}.
The Bias is about 4\% and the resolution is less than 20\%.

Based on the aperture described above, the cosmic ray spectra of proton and H\&He are predicted according to the  H\"{o}randel model~\cite{hrd} and the ARGO-YBJ \& WFCT model~\cite{argo-wfcta} which is an experimental model.
The exposure time is set to one year with 10\% duty cycle.
Since the experimental model ARGO-YBJ \& WFCT only shows the energy spectrum of H\&He, 
the spectrum of pure proton is obtained by dividing by 2, 
that is, the ratio of proton and helium is 1:1.
The bending of the knee keeps unchanged.

\begin{figure}[htb]
	\centering
	\includegraphics[width=0.45\linewidth]{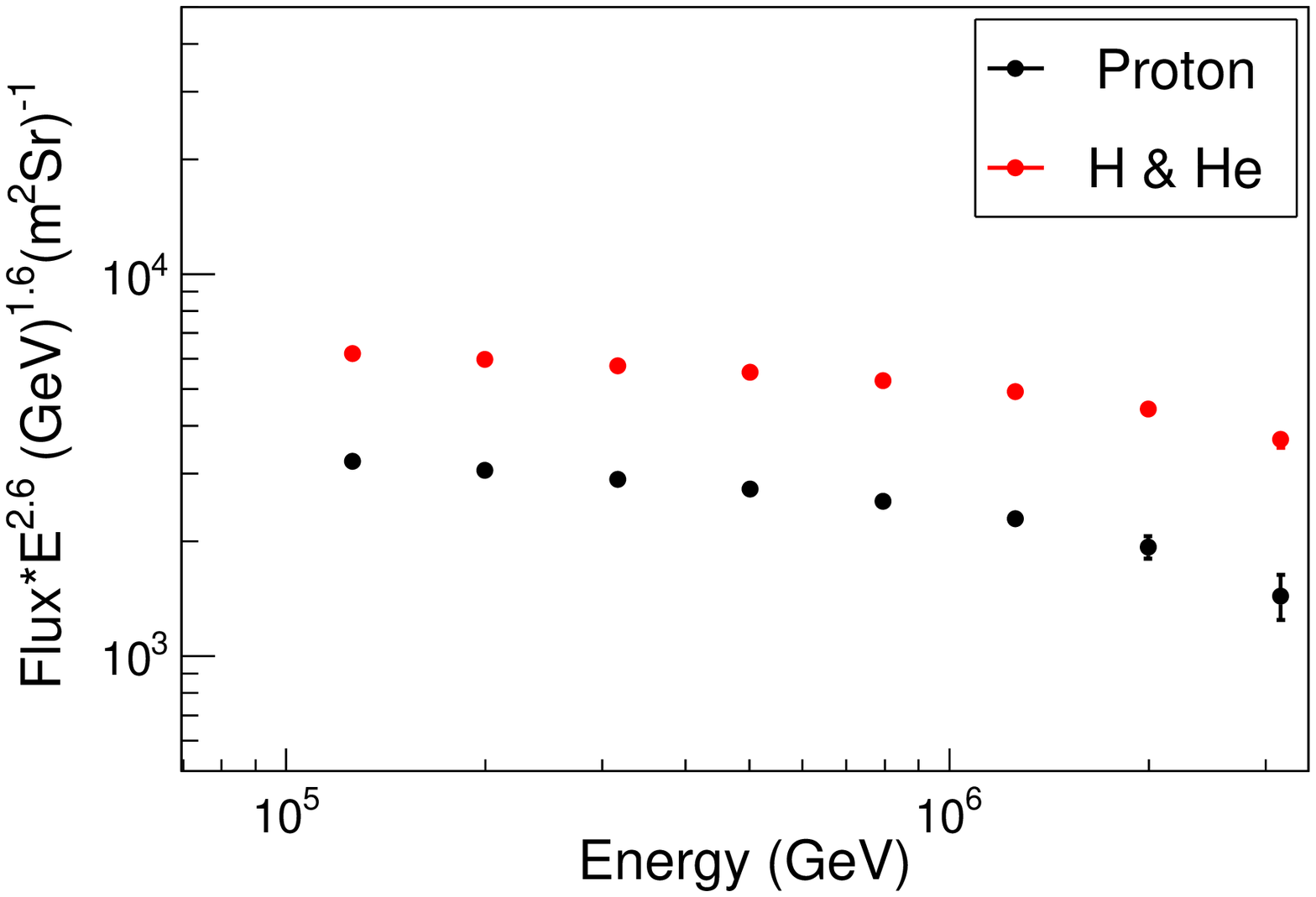}
	\includegraphics[width=0.45\linewidth]{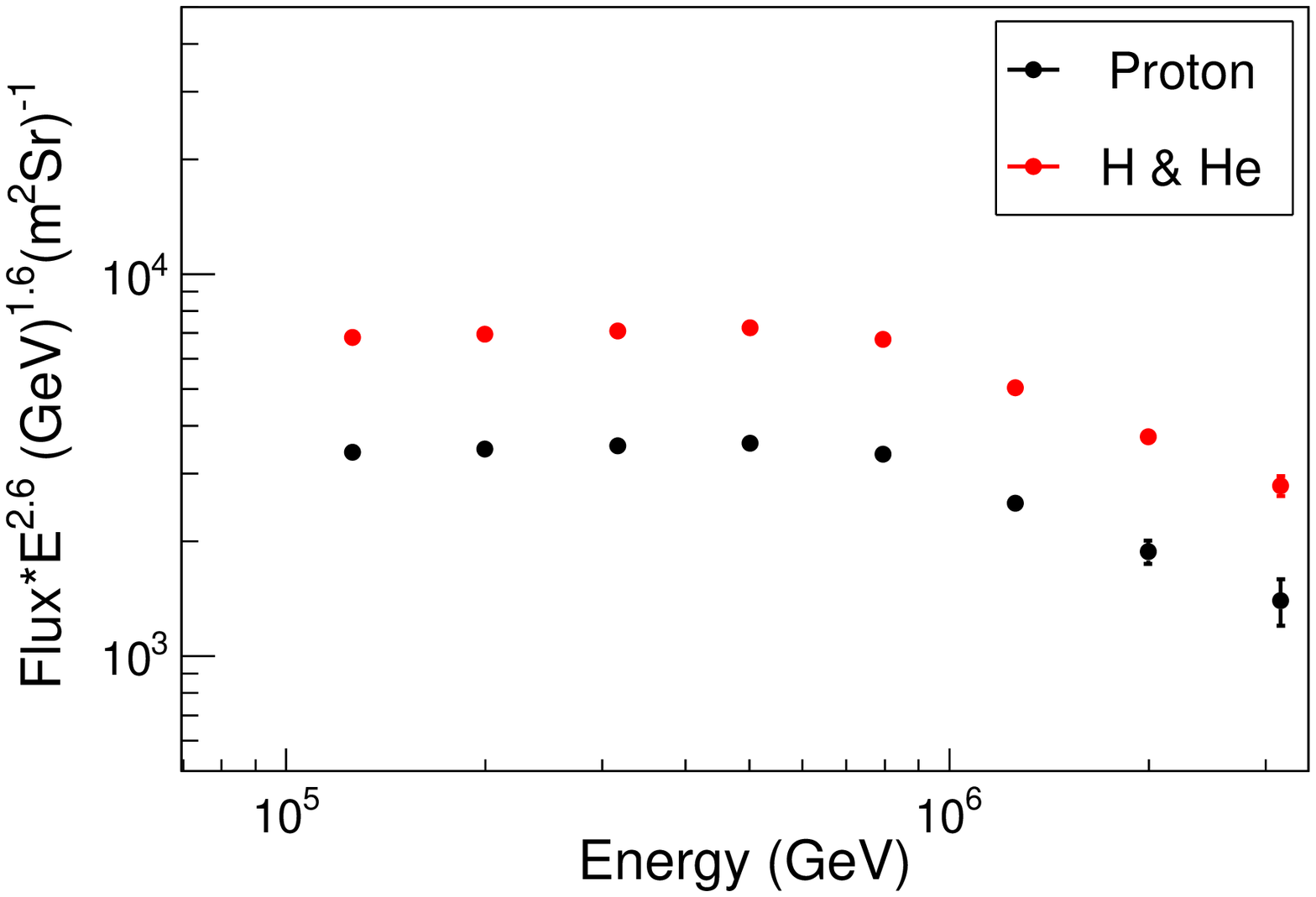}
	\caption{The prospect of spectra at the H\"{o}randel model (left) and the ARGO-YBJ \& WFCT model (right).}
	\label{spectrum-prospect}
\end{figure}

\begin{figure}[htb]
	\centering
	\includegraphics[width=0.45\linewidth]{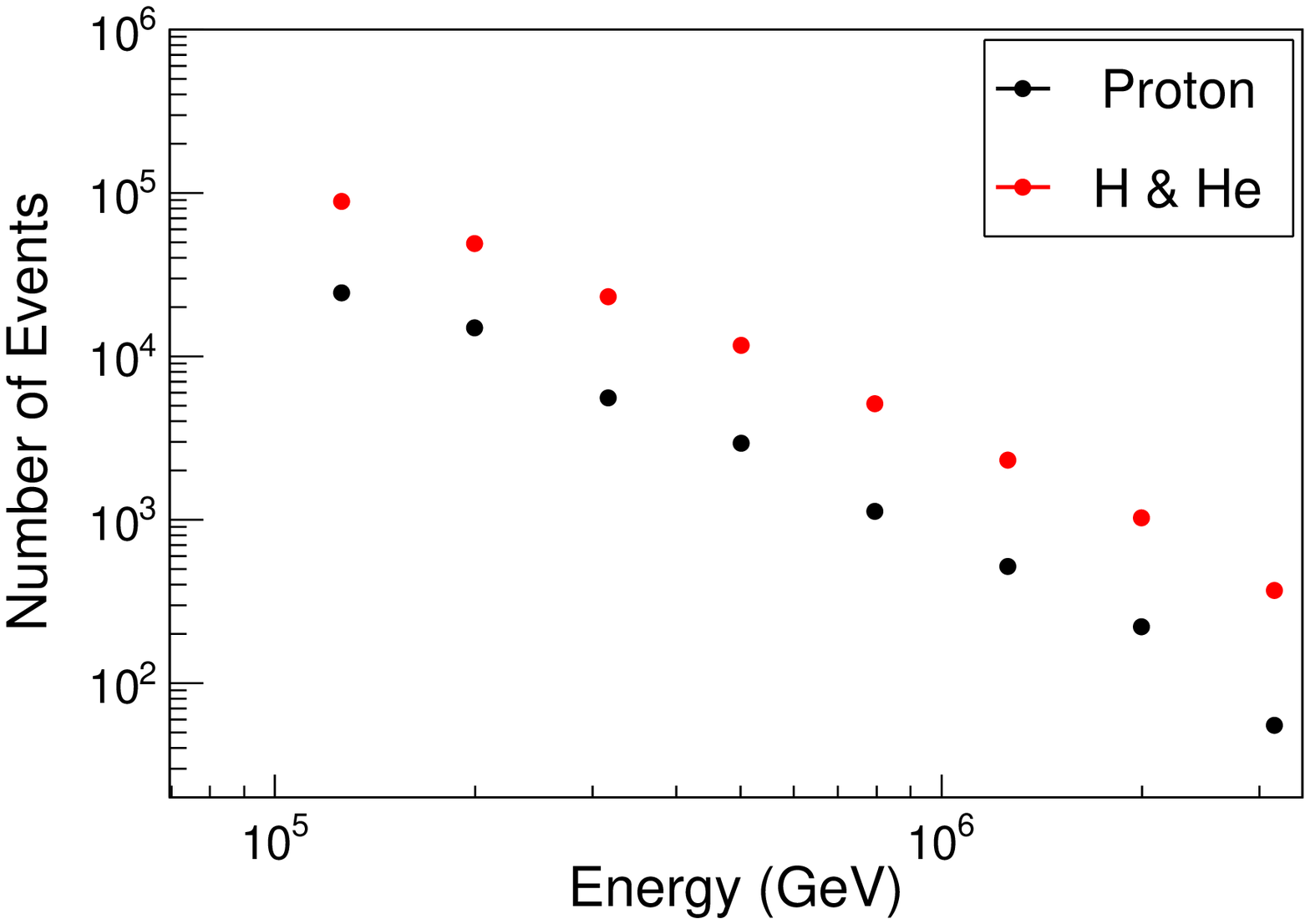}
	\includegraphics[width=0.45\linewidth]{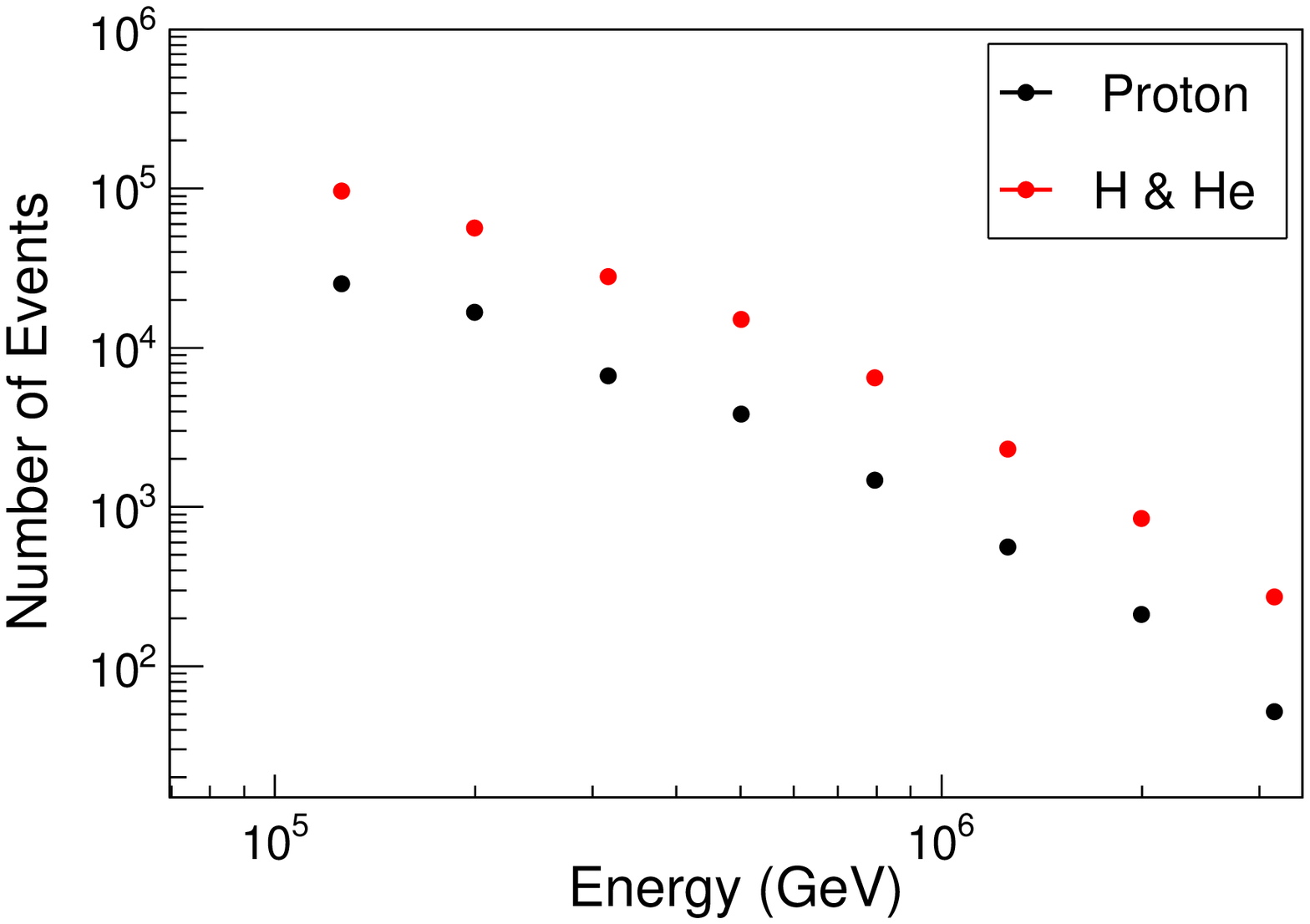}
	\caption{The statistics of one year with 10\% duty cycle at the H\"{o}randel model (left) and the ARGO-YBJ \& WFCT model (right).}
	\label{statistics}
\end{figure}

As shown in Fig.~\ref{spectrum-prospect}, 
the left panel shows the spectrum with the H\"{o}randel model;
and the right panel shows the ARGO-YBJ \& WFCT model.
The black dots show proton and the red dots show $H\&He$.
The statistical error are very small with the aperture and effective time presented above.
The corresponding event rate in one year is shown in Fig.~\ref{statistics}.

At 800 TeV,  5121/6461 H\&He events and 1130/1476 proton events can be selected in one year according to the H\"{o}randel model and the ARGO-YBJ \& WFCT model respectively.
At 2 PeV, 1023/849 H\&He events and 222/210 proton events can be selected in one year according to the H\"{o}randel model and the ARGO-YBJ \& WFCT model respectively.

Therefore, if the knee of the cosmic ray light component is below 1 PeV,
the 1/4 LHAASO array can give good measurements within one year.
If the knee of light component is higher than 3 PeV,
more observation time or more effective methods is needed to get enough statistics.
Considering the application of SiPM, it allows observation in the moon night of WFCTA.
The duty cycle of the hybrid observation can be extended.
Moreover, during the construction of LHAASO, 
more Cherenkov telescopes can also increase the statistics effectively .

\section{Summary}\label{section:summary}

The simulation for the second stage of 1/4 LHAASO hybrid detection is operated.
Three types of detectors, WFCTA, WCDA and MD, are included in this study.
Parameters from each detector array are studied and tuned in detail.
After removing the highly correlated parameters, 
six parameters are used as input of the BDTG classifier for TMVA machine training and testing.
The results show that the classification of pure proton is weaker than that of H\&He.

After cutting, high purity light component samples were selected.
The aperture of pure proton is 900 $m^{2}Sr$, the contamination is around 10\% according to the H\"{o}randel model;
the aperture of H\&He is 1800 $m^{2}Sr$, the contamination is less than 3\%.
For the aperture H\&He, the uncertainty of the different strong interaction models is about $\pm$ 5\%.
Moreover, the bias of energy reconstruction is about $\pm$ 4\%
and the resolution is less than 20\% for both pure proton and H\&He samples.

According to the energy spectrum expectation of the ARGO-YBJ \& WFCT model,
the spectrum of light component in cosmic rays will be measured accurately in a short time by 1/4 LHAASO array.

\section{Acknowledgements}

This work is supported by the National Key R\&D Program of China 
(NO. 2018YFA0404201 and NO. 2018YFA0404202).
This work is also supported by the Key Laboratory of Particle Astrophysics, Institute of High Energy Physics, CAS. 
Projects No. Y5113D005C, No. 11563004 and No. 11775248 of National Natural Science Foundation (NSFC) also provide support to this study.

\end{CJK*}
\end{document}